\documentclass[onecolumn]{aa} 
\usepackage{graphicx,txfonts}
\usepackage[numbers,square,sort&compress]{natbib}

\newcommand{\boomn}{{\sc Boomerang}}
\newcommand{\boom}{{\sc Boomerang} }
\newcommand{\btk}{{\small B03} }
\newcommand{\btkn}{{\small B03}}

\newcommand{\planckhfi}{{\it Planck} HFI }
\newcommand{\planckhfin}{{\it Planck} HFI}
\newcommand{\bicep}{the {\it Robinson} telescope}
\newcommand{\quest}{{\sc Qu}a{\sc d}}
\newcommand{\ebex}{{\sc Ebex}}
\newcommand{\spider}{{\sc Spider}}
\newcommand{\cmb}{{\sc cmb} }
\newcommand{\cmbn}{{\sc cmb}}
\newcommand{\sini}{$\mathrm{Si}_3\mathrm{N}_4$ }

\def\ie{{{\em i.e.}}}

\def\aap{Astr. \& Astroph.\ }

\def\apj{Astrophys.\ J.\ }
\def\apjl{Astrophys.\ J.\ Lett.\ }
\def\apjs{Astrophys.\ J.\ Suppl.\ }

\def\prd{Phys.\ Rev.\ D\ }

\def\gev{GeV~c$^{-2}$}

\def\gevpercm3{\gev~\percm3}
\def\gpercm3{g~\percm3}

\begin{document}

\title{Instrumental and Analytic Methods for Bolometric Polarimetry}

\author{
{W.~C.~Jones\inst{1}} \and
{T.~E.~Montroy\inst{2}} \and
{B.~P.~Crill\inst{3}} \and
{C.~R.~Contaldi\inst{4}} \and
{T.~S.~Kisner\inst{2,5}} \and
{A.~E.~Lange\inst{1}} \and
{C.~J.~MacTavish\inst{6}} \and
{C.~B.~Netterfield\inst{7,8}} \and
{J.~E.~Ruhl\inst{2}}
}

\offprints{William~C.~Jones, wcj@astro.caltech.edu}

\institute{Division of Physics, Math, and Astronomy, 
           California Institute of Technology
\and
           Physics Department, 
           Case Western Reserve University
\and
           Infrared Processing and Analysis Center, 
           California Institute of Technology
\and
           Department of Physics, 
           Imperial College London
\and
           Department of Physics, 
           University of California, Santa Barbara
\and
           Canadian Institute for Theoretical Astrophysics, 
           University of Toronto
\and
           Department of Physics, 
           University of Toronto
\and
           Department of Astronomy and Astrophysics,
           University of Toronto
}

\date{\emph{for submission to ~\aap}}

\abstract
{}{ We discuss instrumental and analytic methods that have been
  developed for the first generation of bolometric cosmic microwave background
  ({\cmbn}) polarimeters.  The design, characterization, and analysis of data
  obtained using Polarization Sensitive Bolometers (PSBs) are described in
  detail.  This is followed by a brief study of the effect 
  of various polarization modulation techniques on the recovery of sky
  polarization from scanning polarimeter data.}
{Having been successfully implemented on the sub-orbital {\boom} experiment,
 PSBs are currently operational in two terrestrial {\cmb} polarization
 experiments ({\quest} and the \emph{Robinson} Telescope).  We investigate
 two approaches to the analysis of data from these experiments, using
 realistic simulations of time ordered data to illustrate the impact of
 instrumental effects on the fidelity of the recovered polarization signal.}
{We find that the analysis of difference time streams takes full advantage
  of the high degree of common mode rejection afforded by the PSB design. In
  addition to the observational efforts currently underway, this discussion is
  directly applicable to the PSBs that constitute the polarized capability of
  the {\planckhfi} instrument.}{} 


\maketitle

\section{Introduction}\label{sec:intro}

Recent advances in millimeter-wave instrumentation and techniques
have transformed observational Cosmic Microwave Background
(\cmbn) research. Over the course of the past decade, statistical
detections of the minute temperature variations in the \cmb have given way to 
high signal to noise imaging of the surface of last scattering. 

The rapid pace of technological development in the field of \cmb research has
contributed to an equally remarkable rate of progress in our understanding of
the Universe; cosmology is in the midst of an abrupt transition from a
data-starved theoretical framework to a rigorously tested standard
model. Remarkably, as of early 2006, the currently avaliable \cmb data are
sufficiently precise (\emph{and accurate!}) that, within the framework of the
most simple inflationary models, the majority of the scientific
potential of the \cmb \emph{temperature} anisotropies has already been
realized.  Measurements of the \emph{polarization} of the \cmb
provide an important confirmation of the validity of the theoretical framework 
through which the data are interpreted, and also aid in the precise
determination of the parameters of the theory.

Further motivation comes from the tantalizing possibility that the \cmb
polarization holds a unique imprint from gravitational waves generated during
the epoch of Inflation. A detection of this signature would represent a
probe of physics beyond the standard model --- both that of cosmology, and
potentially that of particle physics.

Current experimental efforts are focused on the development of high fidelity
\cmb polarimeters capable of characterizing the small fractional polarization
of the \cmbn. The \boomn03 experiment was the first bolometric instrument to
measure the polarization in the \cmbn, and the first of several to use
the Polarization Sensitive Bolometers (PSBs) developed for \planckhfin.  Two
additional terrestrial telescopes using PSBs, Quad and the \emph{Robinson}
Telescope, are currently observing from the South Pole.

We describe the experimental approach employed by the first
generation of bolometric instruments to successfully probe \cmb polarization
and discuss aspects of the data analysis which may inform future observations.
Following a brief description of PSBs, we provide a pedagogical description of
the method of analysis that has been applied to the \btk
data~\cite{b2k_inst,b2k_tt,b2k_te,b2k_ee,b2k_params}. Finally, we investigate the
merits of several modulation schemes for scanning polarimeters which are
directly applicable to current and proposed \cmb polarization experiments at
the South Pole and from Antarctic Long Duration Balloon flights.


\section{Polarization Sensitive Bolometers}\label{sec:psbs}

Quasi--total power and correlation receivers (both
heterodyne~\cite{capmap_inst} and homodyne~\cite{wmap_receiver})
using low noise front-end amplifier blocks based on HEMT amplifiers 
are mature technologies at millimeter wavelengths. The fundamental design
principles of these receivers are well established and have
been used to construct polarimeters at radio to mm-wave
frequencies for many years
~\cite{spiga02,odell_polar,capmap,dasi_pol_inst,cbi_pol}.
Although cryogenic bolometric receivers achieve much higher instantaneous
sensitivities over wider bandwidths than their coherent analogs,
the intrinsic polarization sensitivity of coherent systems has
made them the technology of choice for the first generation of CMB
polarization experiments.

\begin{figure}[t]
\begin{center}
\includegraphics[angle=90,width=15cm]{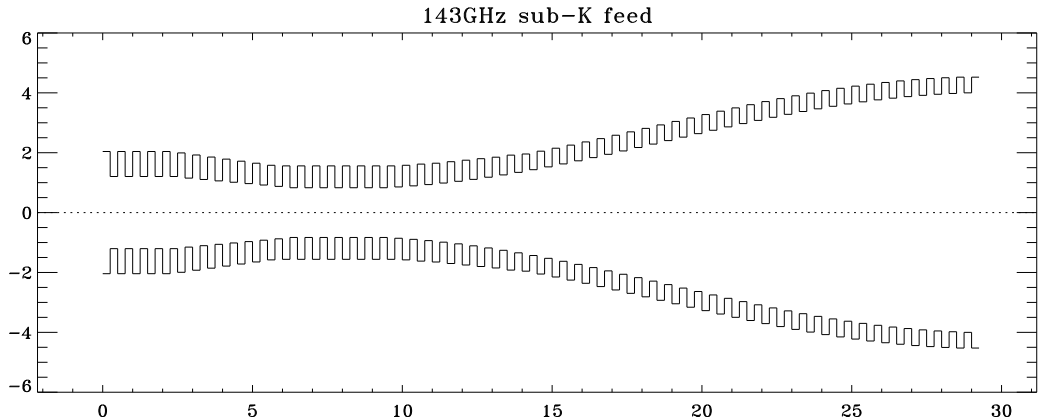}
\end{center}
\caption[Corrugation geometry of the PSB feed]{\small The
  corrugation geometry of the \boom PSB feeds, with dimensions in
  millimeters.  The radiating aperture is on the right hand side, while the
  PSB module is seated on the left.}\label{fig:psbfeed}
\end{figure}

Prior to the release of the \boomn03
results~\cite{b2k_ee,b2k_te,b2k_tt}, all published detections of
\cmb polarization were obtained from experiments relying on the proven HEMT
technology~\cite{dasi_3year,cbi_pol,capmap,wmap_hinshaw,wmap_kogut,wmap_page}.
While interferometry is a robust method of polarimetry, the $N^2$ scaling of
the complexity of correlators prohibits scaling of the design to large optical
throughput, limiting the raw sensitivity of a practical interferometric
experiment. Although much progress has been made in increasing the
sensitivity, and decreasing the footprint, of single-dish HEMT based
correlation receivers~\cite{mmics}, this technology has not yet been
demonstrated with the sensitivity or scalability of contemporary
low--background bolometer arrays at frequencies above $\sim 90$ GHz.

We briefly describe a bolometric system that combines the sensitivity,
stability, and scalability of a cryogenic bolometer with the intrinsic
polarization capability traditionally associated with coherent
systems~\cite{psbs}. In addition, the design obviates the need for
orthogonal mode transducers (OMTs), hybrid tee networks, waveguide plumbing,
or quasi-optical beam splitters whose size and weight make fabrication of
large format arrays impractical. Polarization sensitive bolometers (PSBs) are
fabricated using the proven photolithographic techniques used to produce
``spider web''\ bolometers, and enjoy the same benefits of reduced heat
capacity, negligible cross section to cosmic rays, and structural
rigidity~\cite{yun03}. Finally, unlike OMTs or other waveguide devices,
these systems can be relatively easily scaled to $\sim 600$ GHz,
limited at high frequencies only by the ability to reliably
manufacture sufficiently small single-moded corrugated structures.
Receivers using this design have been demonstrated at 100, 150, 217 
and 353 GHz.

Polarization sensitivity is achieved by controlling the
vector surface current distribution on the absorber, and thus the
efficiency of the ohmic dissipation of incident Poynting flux.
This approach requires that the optics, filtering, and coupling
structure preserve the sense of polarization of the incident
radiation with high fidelity. A multi-stage corrugated feed
structure and coupling cavity has been designed that achieves
polarization sensitivity over a 33\% bandwidth. 

The PSB design has been driven by the desire to minimize
systematic contributions to the polarized signal.  Both senses of
linearly polarized radiation propagate through a single optical
path and filter stack prior to detection, thereby assuring both
detectors have identical spectral passbands and closely matched quantum
efficiencies.

Two orthogonal free-standing lossy grids, separated by $\sim
60 \mu$m and both thermally and electrically isolated, are
impedance-matched to terminate a corrugated waveguide structure.
The physical proximity of the two detectors assures that both
devices operate in identical RF and thermal environments.  A
printed circuit board attached to the module accommodates load
resistors and RF filtering on the leads entering the bolometer cavity.
For \boomn, \planckhfin, QUaD, and the \emph{Robinson} Telescope, the
post-detection electronics consist of a highly stable AC readout with a system
$1/f$ knee below 30 mHz \cite{planckhfi,quad_inst,bicep_inst}.
Unlike coherent systems, this low frequency stability is attained without
phase switching the RF signal. 

We have designed the optical elements, including the
feed antenna and detector assembly, to preserve sky polarization
and minimize instrumental polarization of unpolarized light.  To
this end, the detector has been designed as an integral part of
the optical feed structure.
Corrugated feeds couple radiation from the telescope to
the detector assembly.  Corrugated horns are the favored feed
element for high performance polarized reflector systems due to
their superior beam symmetry, large bandwidth, and low sidelobe
levels. In addition, cylindrical corrugated feeds and waveguides
preserve the orientation of polarized fields with higher fidelity
than do their smooth-walled counterparts.

\begin{figure}[t]
\begin{center}
\includegraphics[bb=190 510 440 700,clip=true,viewport=0 0 250 250,scale=0.9]{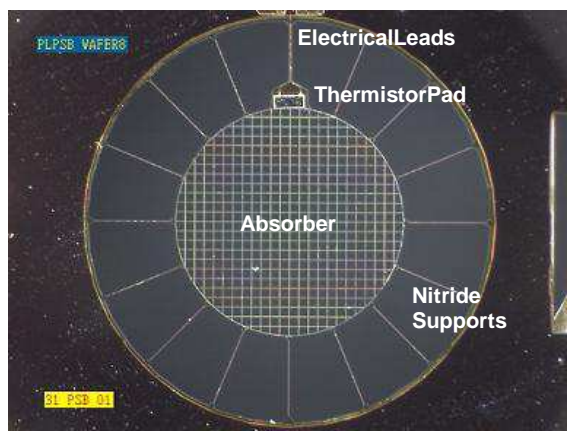}
\end{center}
\caption[Photograph of a PSB]{\small A photograph of a 145 GHz \boom
  PSB absorber.  The diameter of
  the grid is 2.6 mm, while the absorber leg spacing, $g$, is 108
  $\mu$m. Each leg is $3 \mu$m wide.  This device is sensitive to
  incident radiation polarized in the vertical direction due to the
  metalization of the \sini mesh in that direction. The horizontal
  \sini beams evident in the photo are not metalized, and provide
  structural support for the device. The thermal conductivity
  between the absorber and the heat sink is dominated by the
  metallic leads running to the thermistor chip.}
\label{fig:absgeom}
\end{figure}

 The coupling structure, which is cooled to below 1 K,\footnote{The PSB feeds
   in \boomn03, Quad, and the \emph{Robinson} Telescope operate at 240 mK,
   while the \planckhfi focal plane is cooled to $\sim100$ mK.}
consists of a profiled corrugated horn, a modal filter, and an
impedance-matching section that allows efficient coupling to the
polarization sensitive bolometer (see Figure \ref{fig:psbfeed}).
In addition to a reduction in the physical length of the
structure, the profiled horn provides a nearly uniform phase front
that couples well to the other filters and optical elements in
the system. The modal filter isolates the detectors from any
unwanted higher order modes that may be excited in the thermal
break. In addition, this filter completely separates the design
of the bolometer cavity from that of the feed, which couples to the
optics. The impedance-matching section (the re-expansion at the
left side of Figure \ref{fig:psbfeed}) produces a uniform vector
field distribution with a well-defined guide
wavelength\footnote{The guide wavelength, $\lambda_\mathrm{g}$, is
 typically 20\% larger than free space, and
 $\mathrm{d}\log(\lambda_\mathrm{g})/\mathrm{d}\log(\nu)$ 
remains small over the entire range of operation.} and characteristic
impedance over a large ($\sim 33\%$) bandwidth.

The detector assembly is a corrugated waveguide that is operated well above
cutoff. The bolometers, which act as an impedance-matched termination of the
waveguide cavity, are coupled via a weak thermal link to the temperature bath.
The electric field in the cavity drives currents
on the surface of the absorber, resulting in ohmic power
dissipation in the bolometer. This power is detected as a temperature rise
measured by means of matched Neutron Transmutation Doped Germanium
(NTDGe) thermistors ~\cite{beeman}. The bolometers each couple to a single
(mutually orthogonal) linear polarization by precisely matching
the absorber geometry to the vector field of the coupling
structure.  The coupling structure has been tailored to ensure
that the field distribution resulting from a polarized source is highly linear
at the location of the bolometer.

Because the absorber geometry influences the field distribution within the
coupling structure, a treatment of the bolometer cavity as a black-body is in
general {\em not}\, valid. An important consequence of this fact is that any
attempt to model an analogous multimoded optical system must consider
interference terms between modes when calculating coupling efficiencies or
simply trying to predict radiation patterns. The amplitude and phase of any
higher order modes capable of propagating to the bolometer depend on the
details of both the excitation and structure. Therefore, any numerical
calculation would be susceptible to a large number of uncertainties associated
with the appropriate boundary conditions at the bolometer. For this reason, it
may prove difficult to extend the general single mode PSB design
to a multimoded application without sacrificing crosspolar performance.


\section{Analysis}\label{sec:analysis}

\subsection{Polarization formalisms}\label{sec:polform}

The two most commonly used conventions for treating polarized
radiation are the Jones and the Stokes/Mueller formalisms.  
The primary difference between the two approaches is that the
Stokes/Mueller formalism manipulates irradiances, and therefore is
applicable only to incoherent radiation. On the other hand, the Jones
formalism models optical elements with matrix operations on the
(complex) field amplitudes, making it the appropriate approach for
coherent analysis. While the Jones formalism is rather intuitive, the
Stokes formalism is more naturally suited to \cmb analysis.  In the
following we introduce both approaches at an elementary level, and
describe the correspondence between the two.  A more detailed
description of each approach may be found
in~\citet{mueller,jones41a,jones41b,jones42,hecht} and \citet{iauQU}.

The general action of linear optical elements can be described in
terms of the relationship between the input and output electric field vectors.
The Jones matrix of an optical element is defined in terms of its
action on the incident fields,
$$\mathbf{e}_f = \mathbf{J}~\mathbf{e}_i~, $$
where the Jones matrix, $\mathbf{J}$, of the system is a general
product of the matrices describing individual components in the system.
\begin{equation}
\left(\begin{array}{c} E_x \\ E_y \end{array}\right)_f = 
\left[\begin{array}{cc} J_{xx} & J_{xy} \\ J_{yx} & J_{yy} \end{array}\right]_0
\ldots
\left[\begin{array}{cc} J_{xx} & J_{xy} \\ J_{yx} & J_{yy} \end{array}\right]_n
\left(\begin{array}{c} E_x \\ E_y \end{array}\right)_i
\end{equation}
Of course, all such components may be rotated with respect to one
another with the usual rotation matrices,
$$\mathbf{J}^\prime = \mathbf{R~J~R}^T ~,$$
with
$$R \equiv \left(\begin{array}{cc} \cos\psi & -\sin\psi \\
    \sin\psi & \cos\psi \end{array}\right)~.$$
This formalism allows a fairly complicated optical system to be
described by a single matrix, which need only be derived once from the
constituent components.

As an example we describe an imperfect polarizer oriented at an angle
$\psi$ with respect to the basis in which the fields are defined.
Such an object may be represented by the Jones matrix:
\begin{eqnarray}
\mathrm{J}_p &\equiv&
\mathbf{R}~\left[\begin{array}{cc} \eta & 0 \\ 0 & \delta
  \end{array}\right]~\mathbf{R}^T \\
& = & \left[\begin{array}{cc} \eta\cos^2\psi +\delta\sin^2\psi & (\eta-\delta)\cos\psi\sin\psi \\
(\eta-\delta)\cos\psi\sin\psi & \eta\sin^2\psi +
    \delta\cos^2\psi \end{array}\right]~,
\end{eqnarray}
where $\eta > \delta$.  A perfect polarizer would have $\eta =1$ and
$\delta = 0$. After generous application of trigonometric
identities, one recovers the general Jones matrix for an imperfect
polarizer oriented at an angle $\psi$
\begin{equation}
\mathrm{J}_p =
\frac{1}{2}\left[\begin{array}{cc}
    (\eta+\delta)+(\eta-\delta)\cos2\psi  &
    (\eta-\delta)\sin2\psi \\
    (\eta-\delta)\sin2\psi &
    (\eta+\delta)-(\eta-\delta)\cos2\psi\end{array}\right]~.
\label{eqn:polarizer}
\end{equation}
Each detector in a polarization sensitive bolometer pair acts
as just such a partial polarizer, followed by a total power detector.

The Stokes parameters are defined in terms of the electric field as follows:
\begin{center}
\begin{tabular}{ccccc}
$I $ & $\equiv $& $\langle E_x E_x^* + E_y E_y^* \rangle $ &=& $ \langle |E_x|^2\rangle+\langle |E_y|^2\rangle $ \\
$Q $ & $\equiv $& $\langle E_x E_x^* - E_y E_y^* \rangle $ &=& $
  \langle |E_x|^2\rangle -\langle |E_y|^2\rangle $ \\
$U $& $\equiv $& $\langle E_x E_y^* + E_y E_x^* \rangle $ &=& $
  2~\langle |E_x E_y|\cos(\phi_x-\phi_y)\rangle $ \\
$V $& $\equiv $& $ i\langle E_x E_y^* - E_y E_x^* \rangle $ &=& $
  2~\langle |E_x E_y|\sin(\phi_x-\phi_y)\rangle $ \\
\end{tabular}
\end{center}
where the brackets, $\langle~\rangle$, represent a time average and 
the fields are specified in a coordinate system fixed with respect to the
instrument. For Thomson scattering of electrons in a quadrupolar
radiation field there is no mechanism for the introduction of a
relative phase between the two polarizations. Therefore, the
cosmological Stokes $V$ parameter is presumed to be zero.

The action of linear optical elements on a Stokes vector, $\mathbf{s}$, can be
described in terms of the elements\rq ~ Mueller matrix,
$$\mathbf{s}_f = \mathbf{M}~\mathbf{s}_i. $$

Given the definition of the Stokes parameters, one can derive the
relationship between a Jones matrix, and the corresponding Mueller
matrix. Following~\citet{bornandwolf} we find
\begin{equation}
M_{ij}=\frac{1}{2}\mathrm{tr}\left(\mathbf{ \sigma_i J \sigma_j
  J^\dagger}\right)~,
\label{eqn:mueller}
\end{equation}
where the $\sigma_i$ are the Pauli matrices:
\begin{equation}
\begin{array}{cc}
\sigma_I = \sigma_0 \equiv \left(\begin{array}{cc} 
1 & 0 \\
0 & 1 
\end{array}\right)  &
\sigma_Q = \sigma_3 \equiv \left(\begin{array}{cc} 
1 & 0 \\
0 & -1
\end{array}\right) \\ & \\
\sigma_U = \sigma_1 \equiv \left(\begin{array}{cc} 
0 & 1 \\
1 & 0 
\end{array}\right)  &
\sigma_V = \sigma_2 \equiv \left(\begin{array}{cc} 
0 & -i \\
i & 0 
\end{array}\right)~.
\end{array}
\end{equation}

Applying a moderate amount of algebra to Equations \ref{eqn:mueller} and 
\ref{eqn:polarizer}, we find the first row of the Mueller matrix
$\mathbf{M}_p$ for a partial polarizer. This defines the total power
detected as a function of the incident $I$, $Q$, $U$, and $V$ parameters:
\begin{eqnarray}
M_{II} &=& \frac{1}{2}~(\eta^2+\delta^2) \\
M_{IQ} &=& \frac{1}{2}~(\eta^2-\delta^2)~\cos2\psi \\
M_{IU} &=& \frac{1}{2}~(\eta^2-\delta^2)~\sin2\psi \\
M_{IV} &=& 0~.
\end{eqnarray}

The signal from a total power detector is proportional to
the Stokes $I$ parameter of the incident radiation.  Modeling a
polarization sensitive bolometer as a partial polarizer followed by a
total power detector, we find (ignoring, for the moment, the effects
of finite beam size and frequency passband) the data may be expressed
as a sum
\begin{equation}
d_i = \frac{s}{2}\big[ (1+\epsilon)\cdot I + (1-\epsilon) \cdot (
Q\cos 2\psi_i+U\sin 2\psi_i) \big]+n_i~,
\label{eqn:bolosig}
\end{equation}
where we have defined the polarization leakage term, $\epsilon$,
such that $(1-\epsilon)$ is the polarization efficiency\footnote{That
  is, in terms of the elements of the Jones matrix for an imperfect
  polarizer, the leakage $\epsilon \equiv \delta^2/\eta^2$.  This is
  the ratio between the minimum and peak power response to a pure
  linearly polarized source, which is a directly observable property
  of the PSB.}, $\psi$ is the orientation of the axis of sensitivity of the
PSB, and $s$ is the voltage responsivity of the detector. For \boomn, the value
of the crosspolar leakage is typically $\sim5$\%, and less than 3\%
for second generation devices designed for the \planckhfin, BICEP, and QUAD.

It should be noted that the noise contribution, $n$, is overly simplified in
Equation \ref{eqn:bolosig}.  See Appendix \ref{sec:bolonoise} for a more
detailed discussion of the noise properties of bolometric receivers, which
explain the general features of the noise power spectra shown in Figures
\ref{fig:noise} and \ref{fig:sonogram}.

\subsection{Polarized beams}\label{sec:polbeam}

The angular response of an instrument can be characterized by the copolar and 
crosspolar power response functions $P_{\parallel}(r,\theta,\phi)$
and $P_\perp(r,\theta,\phi)$.  In the time reversed sense these can be
thought of as the
normalized power at any point in space resulting from a linearly polarized 
excitation produced by the feed element in the focal plane.  That is, for a 
given polarization $p=(\parallel,\perp)$,
\begin{equation}
P_p(r,\theta,\phi) \equiv
\frac{|E_p(r,\theta,\phi)|^2}{|E_{\parallel}(r,0,0)|^2}~.
\end{equation}

For a single moded system, $P_p$ has nothing to do with the properties of the
detector.  Due to the presence of the modal filter in the throat of the
coupling feed (see Figure \ref{fig:psbfeed}), the beam is a function only of
the feed geometry and the optical elements of the system.   To fully
characterize the system, the polarized beam patterns must be considered
separately from the detector.


The exact definitions used for the polarizations on the sphere
$p=(\parallel,\perp)$ vary in the literature, but the standard is
Ludwig\rq s Third definition \cite{ludwig}. In any case, for small
angles away from the beam centroid, they are very nearly equal to the
Cartesian definition.

Qualitatively, $P_{\parallel}$ is similar to an Airy pattern; a Gaussian
near the beam centroid, with a series of side-lobes.  $P_{\perp}$ is
zero on-axis for most optical systems and, for on-axis systems, is
also zero along the E- and H-planes.  The peak of the crosspolar 
pattern is typically at the
half power point of the copolar beam, and the peak amplitude relative
to the copolar beam is fundamentally related to the \emph{asymmetry} of
the copolar beam.  
For an azimuthally symmetric system, such as a feedhorn antenna, this
produces lobes in the $45^\circ$ plane in all four quadrants of
the beam. For an off-axis reflector such as the \boom telescope, the azimuthal
symmetry is lost and the lobes are bimodal. It is worth noting that
the polarized beams generally depend on frequency as well as the
field distance.

A PSB detects the convolution of the polarized sky with the polarized
  beam, integrated over the frequency bandpass, and subject to the
  polarization efficiency of the detector.\footnote{This treatment is
  actually more general; it holds for any receiver that can be
  characterized as a total power detector preceded by an imperfect
  polarizer.  That is, any receiver that is well described by a Jones
  matrix of the type
$$ J_p=\left[\begin{array}{cc} \eta & 0 \\ 0 & \delta
    \end{array}\right].$$
In the discussion that follows, keep in mind that
$\epsilon\equiv\frac{\delta^2}{\eta^2}$.}
In the flat sky approximation, a time domain sample of a single
detector within a PSB pair, $d_i$, may therefore be written as the sum
of a signal component
\begin{equation}
d_i=\frac{s}{2}\int d\nu \frac{\lambda^2}{\Omega_b} F_\nu\iint
d\Omega~\Big(P_\parallel(\hat{r}_i)+P_\perp(\hat{r}_i)\Big)
\Big[~I+\gamma~\mathcal{P}(\hat{r}_i)~\Big(Q \cos 2\psi_i + U \sin
  2\psi_i~\Big)\Big],
\label{eqn:gensig2}
\end{equation}
and a noise contribution.  Here, the Stokes parameters are defined on
the full sky and the integration variable is $\hat{r}_i =
\hat{n}_i-\hat{r}$, for a vector, $\hat{n}_i$, describing the pointing
at a time sample, $i$. We have also defined the beam solid angle
$\Omega_b = \iint d\Omega~(P_\parallel+P_\perp)$. The normalized beam
response and the polarization efficiency are given by,
\begin{equation}
\begin{array}{ccc}
\mathcal{P}(\hat{r}) \equiv \frac{P_{\parallel}-P_\perp}{P_{\parallel}+P_\perp}
& \hspace{5mm} &
\gamma \equiv \frac{1-\epsilon}{1+\epsilon}
\end{array}.
\label{eqn:tod}
\end{equation}
For B03, the angle $\psi$ is modulated by sky rotation and the motion of the
gondola.  The calibration factor, $s$, converts the brightness
fluctuations in $I$, $Q$, and $U$ to a signal voltage.

By rearranging Equation \ref{eqn:gensig2} and dropping both the explicit
spatial and frequency dependencies, the relation can be written more
intuitively, 
\begin{equation}
d_i \simeq \frac{s}{2}\int d\nu~\lambda^2~F_\nu\iint
d\Omega~\left[~I
  +~\gamma~\mathcal{P}~\left(~Q \cos 2\psi_i + ~U \sin 2\psi_i\right) \right].
\label{eqn:gensig}
\end{equation}
We have made the simplifying assumption that we may remove the
beam and polarization efficiencies from the integral over the sky, and
then absorb these prefactors into a redefinition of the calibration constant,
$s = s^\prime \int d\nu~(1+\epsilon)$.


\subsection{Signal and Noise Estimation}\label{sec:sig}

A great deal of effort has been devoted to the development of algorithms
designed to estimate the signal and noise from noise-dominated data, and a
rich literature has developed around the topic (for some recent examples,
see \citet{wmap_3yr_mapmaking,dore_map,amblard}). In this section
we outline in pedagogical detail the method used to estimate the signal and
noise from the published \boomn03 data~\cite{b2k_params,b2k_ee,b2k_te,b2k_tt}.

An estimate of the instrumental noise properties that is both precise
and accurate is required in order to avoid the introduction of a bias
to the estimate of the power spectrum of the signal.  For \btkn, the high
signal to noise ratio of the data in fact complicates the noise estimation
procedure. We solve for the noise and signal simultaneously using an
iterative procedure adapted from that applied in the analysis of the
data from the 1998 flight of \boomn~\cite{prunet,netal,retal}.
The B03 data are unique among contemporary \cmb experiments in that not
only are the temperature maps signal dominated at angular scales approaching
the beamsize, but the time ordered data are
also characterized by signal to noise ratios of order unity. \btk
is the only polarized dataset that is comparable in this regard to that
anticipated from the \planckhfin. 

Assuming that the data, $\mathbf{d}$, are well described as the sum of a
sky signal and a noise contribution,
$\mathbf{d}= \mathbf{Am}+\mathbf{n}$,
where $\mathbf{A}^T$ is the pointing matrix which maps time domain
samples to pixels on the sky.  If the statistical properties of the
noise contribution are piecewise stationary with a (circulant)
noise covariance matrix, $\mathbf{N}$, defined as
$$N_{tt^\prime} = \frac{1}{N}\sum_i^{N_d} \left(n_{i+t} -
  \langle\mathbf{n}\rangle\right) \left(n_{i+t^\prime}-
  \langle\mathbf{n}\rangle \right),$$
then the least squares estimate of the map is given by,
\begin{equation}
\widetilde{\mathbf{m}} =
\left(\mathbf{A}^T\mathbf{N}^{-1}\mathbf{A}\right)^{-1}\mathbf{A}^T\mathbf{N}^{-1}\mathbf{d}.\label{eqn:map}
\end{equation}
The iterative procedure begins with the assumption of a white noise power
spectrum (\ie, diagonal $\mathbf{N}$), in which case Equation
\ref{eqn:map} corresponds to a simple average of the data falling in a 
given pixel.  The noise contribution used in a given iteration on the
solution to Equation \ref{eqn:map} is obtained from the estimate of
the signal obtained in the previous iteration,
\begin{equation}
\widetilde{\mathbf{n}}_{k+1} = \mathbf{d}-\mathbf{A}~\widetilde{\mathbf{m}}_k.
\label{eqn:noise}
\end{equation}
Piecewise stationarity of the noise allows subsequent convolutions of
$\mathbf{N}^{-1}\mathbf{d}$ to be performed in the Fourier domain.

For most terrestrial telescopes, which suffer from relatively high backgrounds
and large atmospheric signals, the time stream is dominated by
noise.  Therefore, a good approximation to the noise covariance matrix
appearing in Equation \ref{eqn:map} is provided by the power spectrum of the
raw data. For orbital and balloon-based \cmb experiments like \boomn, the
time ordered data are not noise dominated, greatly
complicating an accurate determination of the noise. In general, an
in-situ estimation of the noise is required due to the influence of
atmospheric emission, unpredictable backgrounds, and scan-synchronous effects.
As a result, the simultaneous estimation of both the signal and noise
is required.

\begin{figure}[!t]
\begin{center}
\rotatebox{90}{\scalebox{0.6}{\includegraphics{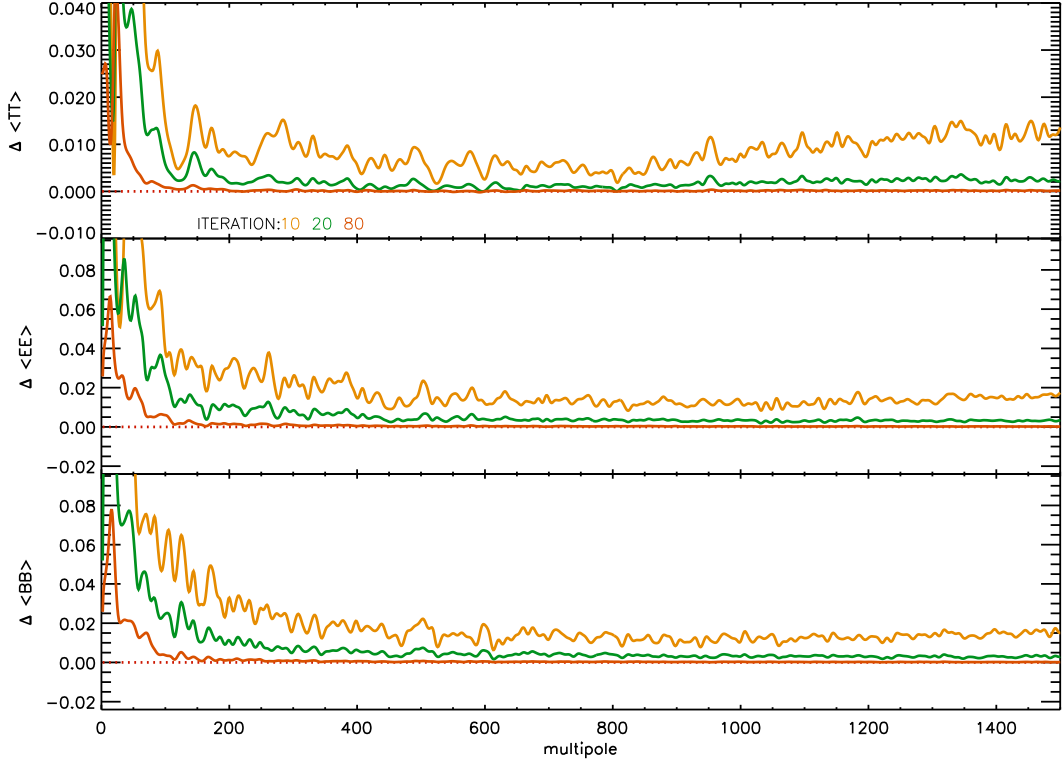}}}
\caption[Convergence of iterative GLS mapmaker]{\small The
  \emph{fractional change} in the cut-sky angular power spectra
  (pseudo-$C_\ell$) for temperature (\emph{tom panel}), and polarization
  (\emph{bottom panels}) from iteration to iteration of the mapmaker. The 
  largest scales take the longest to converge, and polarization takes longer
  to converge than does temperature.  Note that the scale of the polarization
  axes are increased by a factor of two relative to the temperature.}
\label{fig:converge2}
\end{center}
\end{figure}

An iterative solution for both $\mathbf{N}$ and $\mathbf{\widetilde{m}}$
is possible by using an adaptation of the Jacobi method.  The Jacobi
method is an iterative approach to the solution of a general linear system of
equations, such as Equation \ref{eqn:map}, that does not require the
inversion of large matrices.  The application to the solution of
Equation \ref{eqn:map} is derived in Appendix \ref{sec:jacobi}.  This
algorithm is naturally suited to the problem of noise estimation, as
the signal subtraction is an integral part of the iterative
procedure.

By iterating on the noise covariance
matrix, $\mathbf{N}$, as well as the signal, $\mathbf{\widetilde{m}}$, one
approaches a general least squares solution for both.
This procedure has been used in the noise estimation of previous
experiments that probed the \cmb temperature
anisotropies~\cite{dmrmaps,prunet}. In this application, the approach
has been extended to a polarized data set.

\begin{figure}[!t]
\begin{center}
\rotatebox{90}{\scalebox{0.28}{\includegraphics{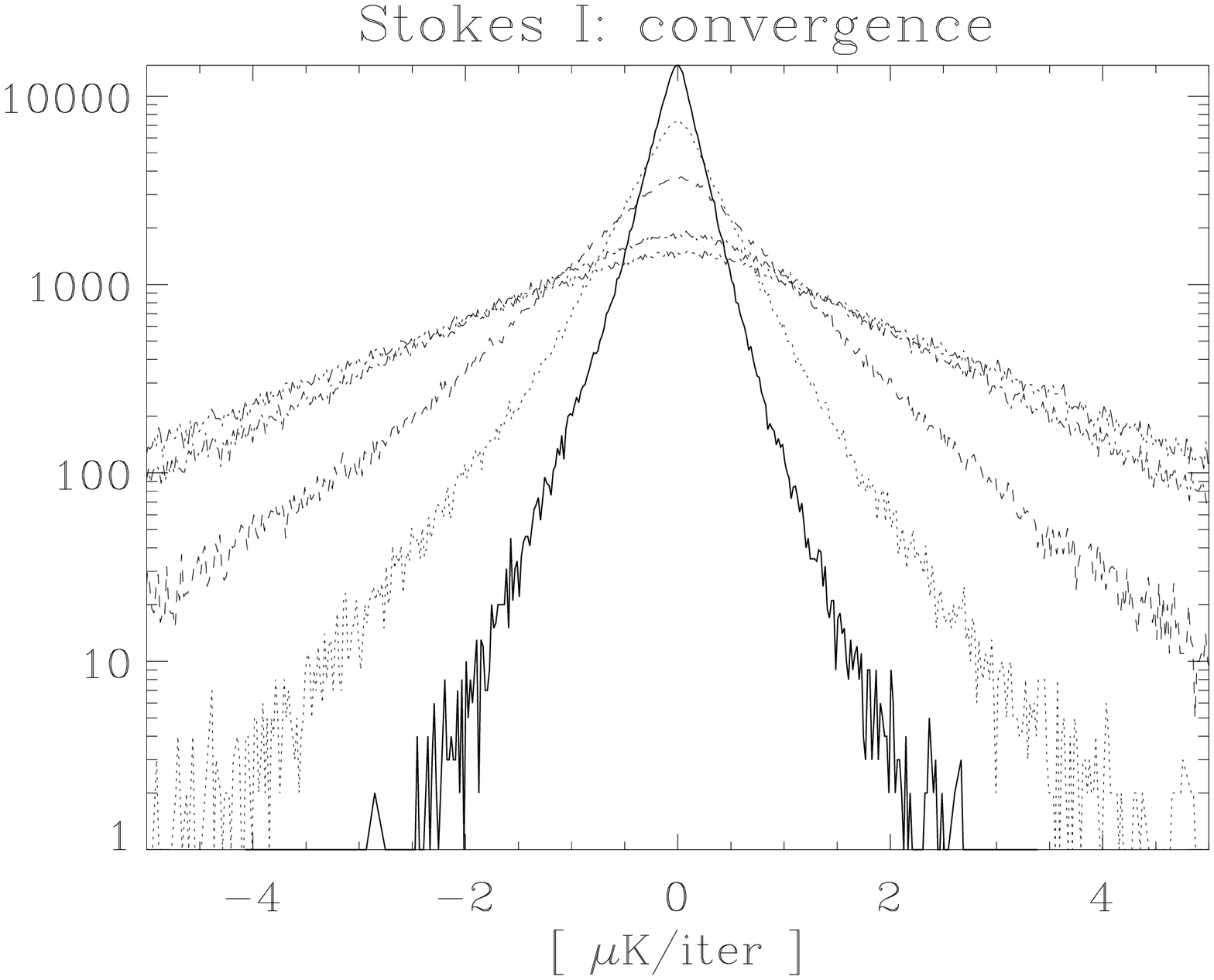}}}
\rotatebox{90}{\scalebox{0.28}{\includegraphics{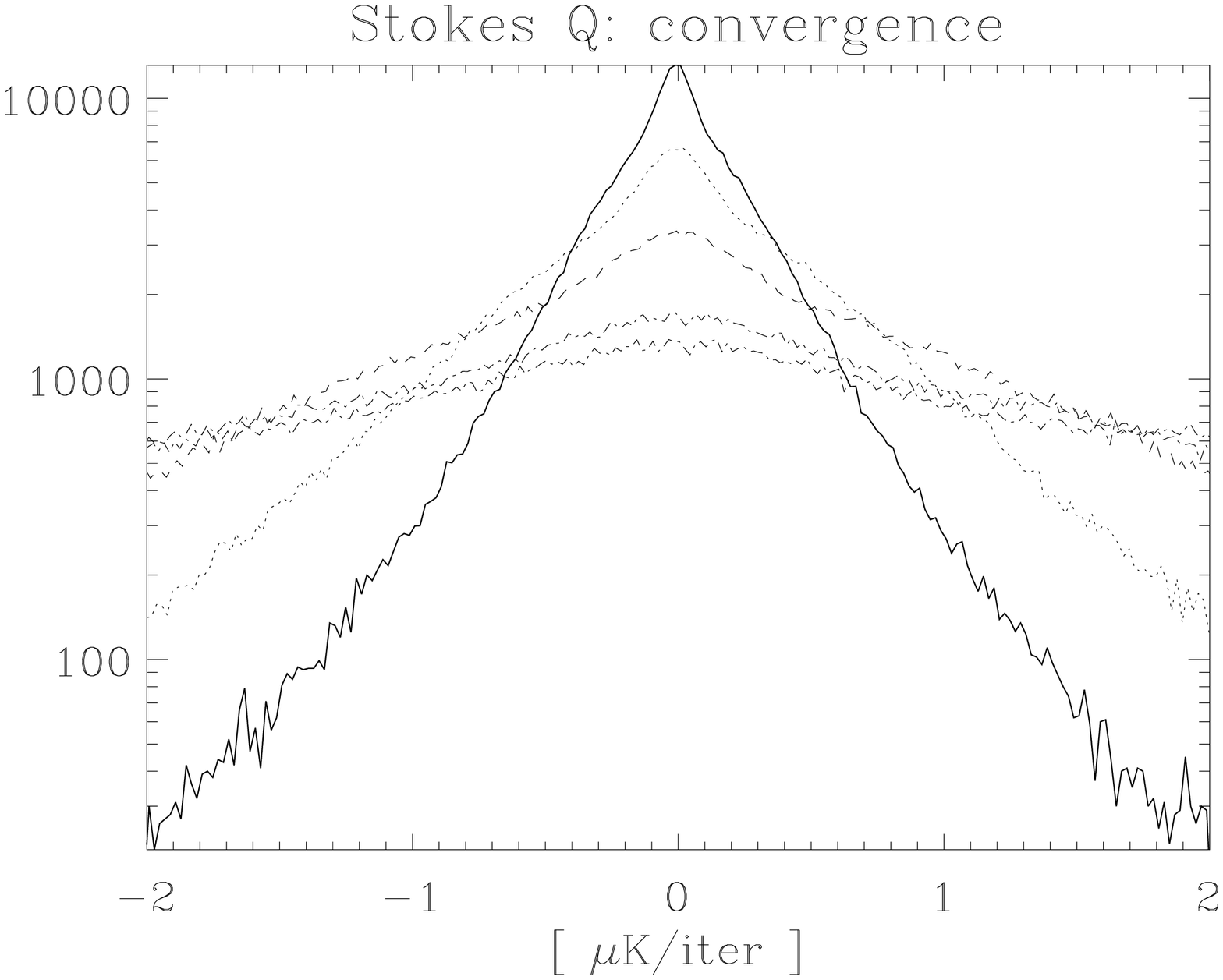}}}
\caption[Convergence of iterative GLS mapmaker]{\small The
  convergence criteria for the iterative procedure is determined by a
  threshold on the {\it rms} amplitude of the correction.  Histograms of
  the corrections to the $I$ \emph{(left panel)} and $Q$ \emph{(right panel)}
  maps (the change in pixel value between subsequent iterations) are shown for
  5, 10, 20, 40 and 80 iterations.}
\label{fig:converge3}
\end{center}
\end{figure}

As described in Appendix \ref{sec:jacobi}, each subsequent
iteration on the solution to Equation \ref{eqn:map},
$\mathbf{\widetilde{m}}_{k+1}$, is calculated from the previous solution
according to the procedure
$$ \mathbf{\widetilde{m}}_{k+1}=\mathbf{\widetilde{m}}_k+\delta
\mathbf{\widetilde{m}}_{k+1}~, $$
where
$$ \delta \mathbf{\widetilde{m}}_{k+1} \equiv
~\alpha~\cdot~\mathrm{diag}(\mathbf{A}^T \mathbf{N}^{-1}_k
\mathbf{A})^{-1} \mathbf{A}^T \mathbf{N}^{-1}_k (\mathbf{d}-\mathbf{A}
~\mathbf{\widetilde{m}}_k)~, $$
and the relaxation parameter, $\alpha \lesssim 1$, is tuned to
optimize the speed of convergence. 
Recall that, in the case of polarized data, the quantity $\mathbf{d}$
represents the left hand side of Equation \ref{eqn:decor}.  Therefore,
the calculation of the matrix $\mathrm{diag}(\mathbf{A}^T
\mathbf{N}^{-1}_k \mathbf{A})^{-1}$ involves the inversion of the
polarization decorrelation matrix on the right hand side of Equation
\ref{eqn:decor}.
The great advantage of this method is that the convolution of the data
with the inverse noise correlation matrix,
\begin{equation}
\mathbf{n}_{k+1} \equiv
\mathbf{N}_{k}^{-1}(\mathbf{d}-\mathbf{A}~\widetilde{\mathbf{m}}_k)~,
\label{eqn:noisefilter}
\end{equation}
can be efficiently calculated in the Fourier domain, without needing to
invert the full time domain correlation matrix, $\mathbf{N}$.
This operation is simply the application of a Fourier filter to the
signal-subtracted time stream using the inverse of the noise power spectrum as
the filter kernel. 

The unbiased estimation of power spectra relies crucially on
the ability to accurately model the noise properties of the
instrument~\cite{master,madcap1}.  In order to treat the noise in a
self-consistent fashion as the realization of a Gaussian random process,
it is necessary to measure and store the observed auto- \emph{and}
cross-correlations for all channel permutations, and each noise stationary
subset. In the North American analysis of \btkn,\footnote{The \boom team
  implemented two independent analysis of the time ordered data from the 2003
  Antarctic Long Duration Balloon flight.  The results from both pipelines are
  reported in \citet{b2k_tt,b2k_ee,b2k_te,b2k_inst}.} the data are divided into 215
noise-stationary subsets, referred to as chunks, each of which consist of
approximately one hour of data. For each of these chunks
the 36 (complex) auto- and cross-power spectra are calculated, binned
logarithmically, and stored to disk.  When used to generate noise realizations
or construct filtering kernels, these binned spectra are interpolated to the
discrete frequencies required by each subset of the data.

\begin{figure}[!t]
\begin{center}
\rotatebox{90}{\scalebox{0.7}{\includegraphics{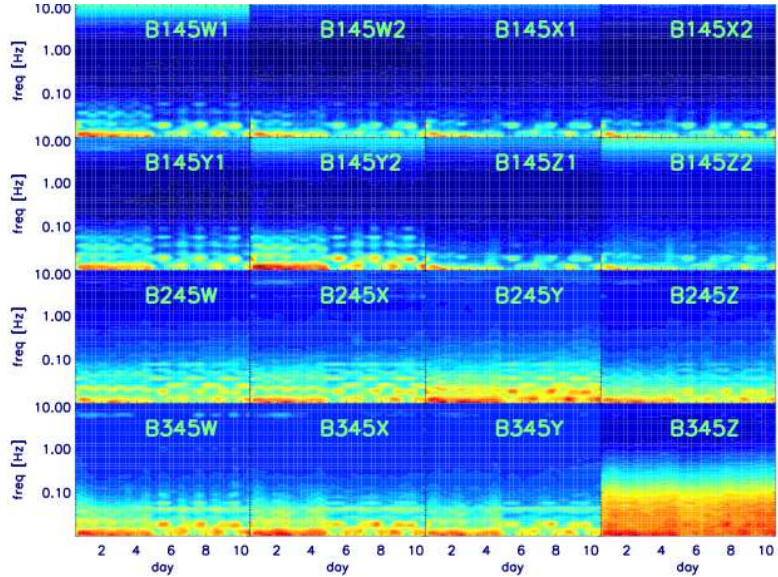}}}
\caption[Noise sonogram]{\small The time dependence of the (signal
  subtracted) noise power spectra of the \boom science channels as
  determined from the in-flight data.  Each frame shows the power
  spectrum of each noise stationary subset (chunk) from a particular
  channel.  The series of lines above 10 mHz corresponds to the
  harmonics of the scan frequency.  The signal band extends from 0.05
  to 5 Hz.  The diurnal dependence of the $1/f$ knee is evident.  The
  B345Z channel exhibited noise whose properties were neither
  stationary nor Gaussian, which is manifest in the low frequency
  contribution.}
\label{fig:sonogram}
\end{center}
\end{figure}

The \boom readout electronics are AC coupled
at $\sim 6$ mHz, and therefore there is no useful information in the
time stream on timescales longer than that set by the stationarity of the
noise. Dividing the data into these hour long subsets represents a tradeoff 
between sample variance and stationarity in the accuracy of the noise
estimate.  The non-stationarity of the \btk noise is illustrated in
Figure \ref{fig:sonogram}.

The chunk boundaries are chosen to maximize the accuracy of the noise estimate.
The length of these chunks introduces a practical limit to the length,
$N_{\tau}$, of the kernel applied in Equation \ref{eqn:noisefilter}.  The
computational scaling is thus $N_d \log(N_\tau)$ for each iteration of the
mapmaker.  The memory requirement is also set by the degree of noise
stationarity; the algorithm only requires the pointing and bolometer data for
an individual chunk to be held in memory at any given time.  For \btkn, the
contributions of file writing and Fourier transforms to the run time are
approximately equal, depending on the avaliable memory.\footnote{The Jacobi
  solver implemented by the North American \boom team requires $\sim 120$ MB
  of RAM and produces a converged GLS estimate of the signal and noise at a
  rate of 10 processor-$s$/channel/hour of data sampled at 60 Hz, when running
  on a 2 GHz AMD Athlon64 X2 workstation.}

The Fourier approach to the analysis requires the data within a chunk to be
continuous and well characterized by a given noise power spectrum.  About 7\%
of the \btk time stream is contaminated by transient events (primarily cosmic
ray hits and calibration lamp pulses).  These gaps are flagged, and
replaced with fake data that are statistically consistent with the remainder
of the chunk.  The signal subtracted data are easily filled with any
reasonable realization of the noise. Due to the small fraction of the data 
which are contaminated, the exact method of gap-filling has negligible impact
on the final signal and noise estimates.

For each chunk all $\left[N_{ch} (N_{ch}-1)/2 + N_{ch}\right]$ auto and cross
power spectra are derived from the signal-subtracted time stream,
$\mathbf{n}$, obtained from Equation \ref{eqn:map} using the
maximum likelihood maps derived from the full set of data.
The noise spectra obtained in this manner are
generally biased due to the effect of pixelizing the (continuous) sky signal,
as well as the finite signal to noise with which the
sky signal, $\widetilde{\mathbf{m}}$, is recovered~\cite{amblard}.

Given a sufficiently high resolution pixelization, the signal variation within
a pixel can be made to be negligibly small compared to the noise in the
map. We pixelize the sky using the {\sc healp}ix method, at a resolution which
corresponds to a pixel size of $\simeq 3.4^\prime$~\cite{healpix}.  At this
resolution we find the effect of pixelization to be well below the
instrumental noise per pixel of the \btk data.  As described in
\citet{b2k_inst} and \cite{b2k_tt}, the B03 \cmb data are divided into a
shallow and deep field, the latter being a subset of the former.  The noise
per pixel of the deep field is roughly three times lower than the shallow
field. 

The impact of the noise in the signal estimate \emph{is}
found to be significant for the data that constitute the shallow region of
the \btk target field. The raw sensitivity of the instrument ultimately
determines the signal to noise ratio of the time ordered data and,
when combined with the distribution of integration time on the sky,
the fidelity of the recovered signal estimate, $\widetilde{\mathbf{m}}$.
The error in the signal estimate, $\widetilde{\mathbf{m}}$, introduces 
a bias to the estimate of the noise power spectrum~\cite{amblard}. 
This bias is generally frequency dependent because of the finite
bandwidth of the signal.  The bias in the noise estimation varies
from chunk to chunk as a result of the variation in signal to noise
ratio in different parts of the map.

The origin of this bias can be understood through closer examination of the
signal-subtracted time stream, $\widetilde{\mathbf{n}}$, that is obtained from
the estimate of the Stokes parameter maps, $\widetilde{\mathbf{m}}$, namely, 
$\widetilde{\mathbf{n}} = \mathbf{d}-\mathbf{A}\widetilde{\mathbf{m}}$.
The data are assumed to consist of the sum of a pure signal and noise,
$\mathbf{d}=\mathbf{s}+\mathbf{n}$, giving
\begin{eqnarray}
\widetilde{\mathbf{n}}
&=& \mathbf{s}+\mathbf{n}-\mathbf{A}\widetilde{\mathbf{m}}\nonumber \\
&=& \mathbf{n}-\mathbf{\hat{n}} \label{eqn:nbias}
\end{eqnarray}
where we have defined the projection of the signal error to the
time stream as
$\mathbf{\hat{n}}\equiv\mathbf{A}(\widetilde{\mathbf{m}}-\mathbf{m})$.
The raw noise power spectrum,
$\langle\widetilde{\mathbf{n}}\widetilde{\mathbf{n}}^\dagger\rangle$, 
which is estimated from Equation \ref{eqn:nbias} differs from the true noise
power spectrum, $\langle\mathbf{n}\mathbf{n}^\dagger\rangle$, by the factor
\begin{equation}
\left(~1+\frac{\langle\mathbf{\hat{n}}\mathbf{\hat{n}}^\dagger\rangle}{
\langle\mathbf{n}\mathbf{n}^\dagger\rangle}
-2~\frac{\langle\mathbf{\hat{n}n^\dagger}\rangle}{
  \langle\mathbf{n}\mathbf{n}^\dagger\rangle}~\right)
\label{eqn:nbias2}
\end{equation}
For \btkn, the projection of the map errors to the time domain is
highly correlated with the true time domain noise, and therefore the
cross-correlation term dominates in Equation \ref{eqn:nbias2}.  The
raw noise power spectra therefore tend to \emph{underestimate} the
true amplitude of the noise at frequencies within the signal
bandwidth.  The amplitude of the bias term in Equation
\ref{eqn:nbias2} is as high as 10\% for the most poorly covered
regions in the shallow field, and is below 1\% for the 175 chunks of
the deep field.

\begin{figure}[!t]
\begin{center}
\rotatebox{90}{\scalebox{0.3}{\includegraphics{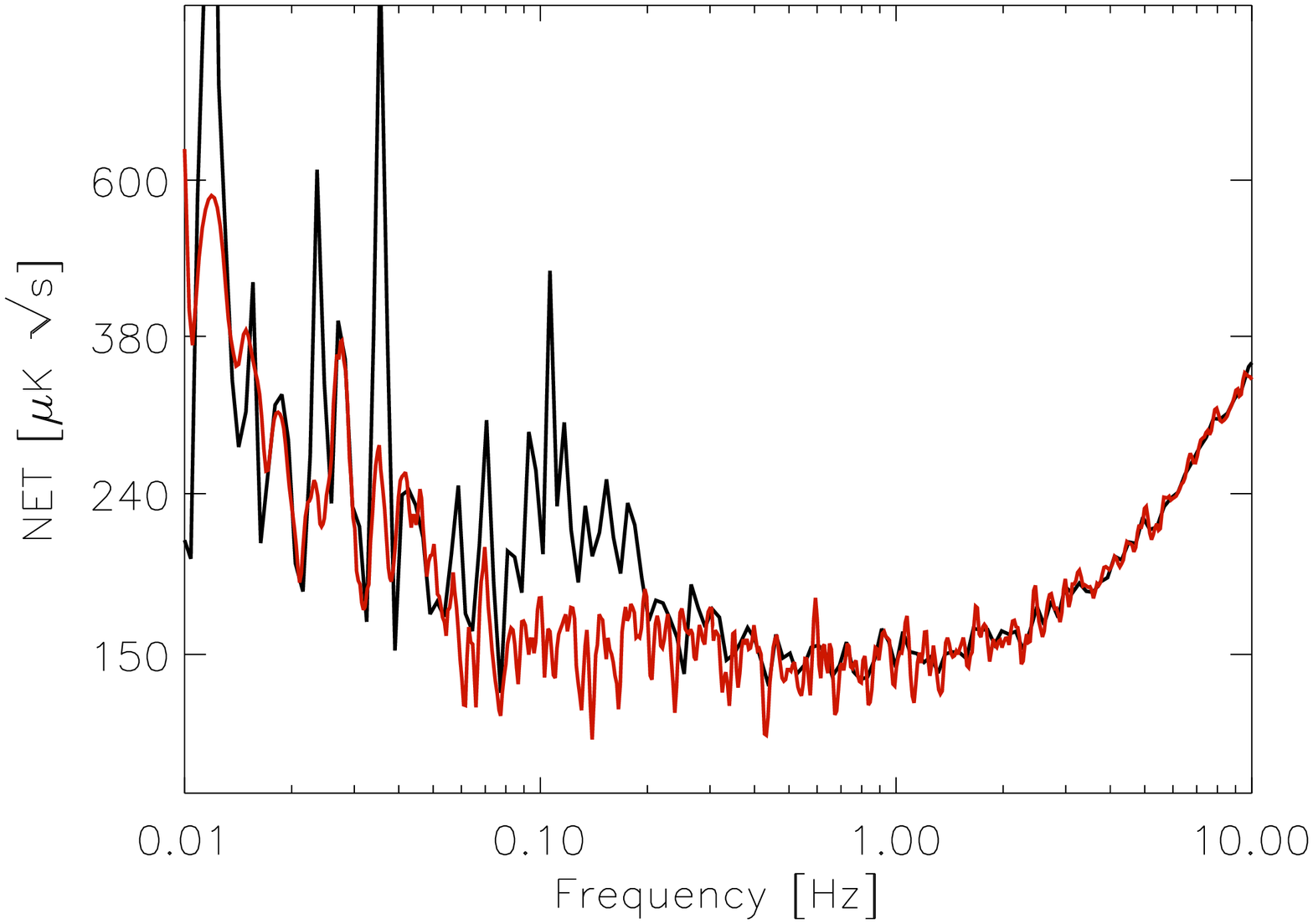}}}
\rotatebox{90}{\scalebox{0.3}{\includegraphics{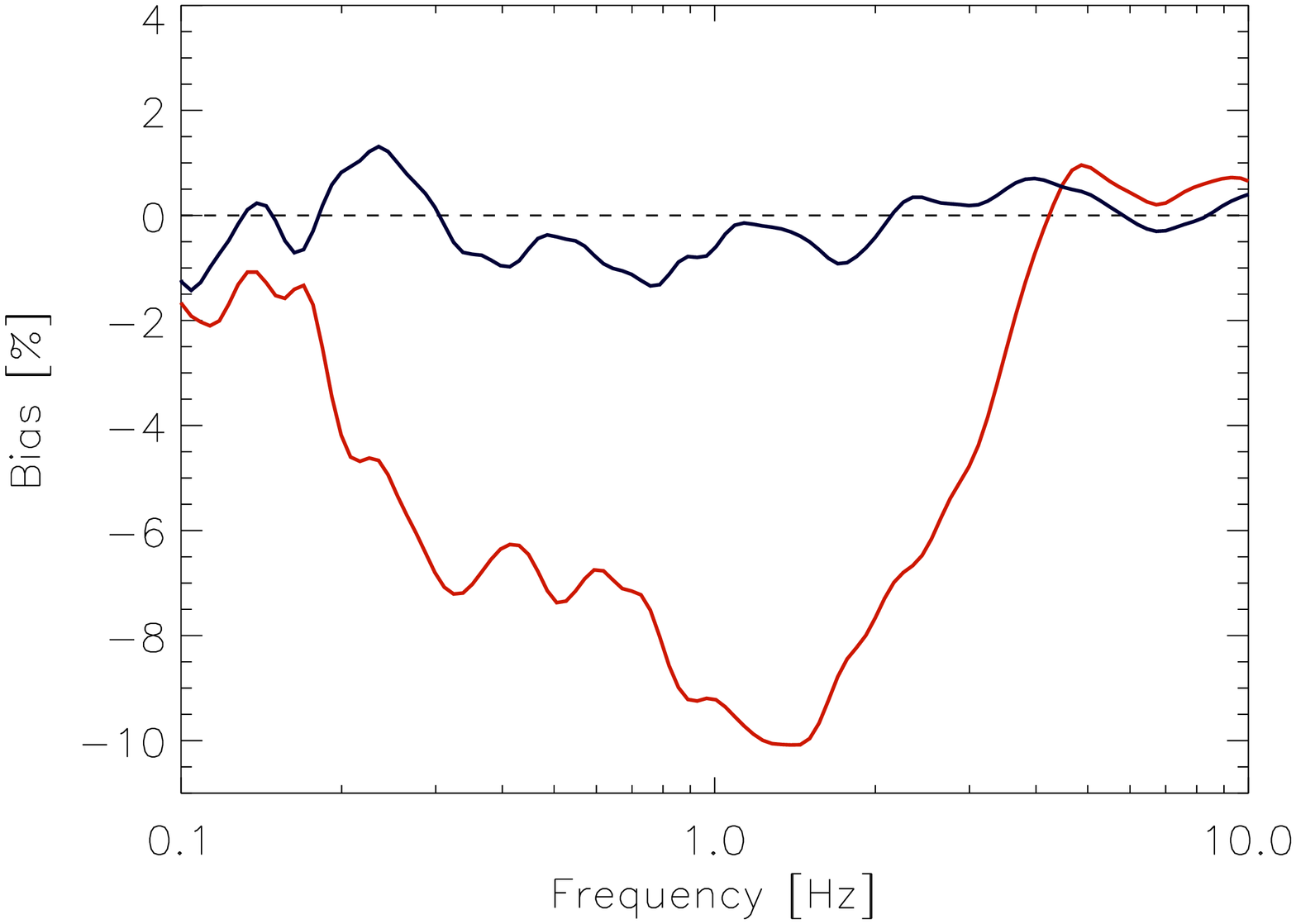}}}
\caption[Noise PSD and bias]{\small \emph{Left panel:} the power
  spectral density, in \cmb units, of the noise for a representative
  chunk of the deconvolved B03 time ordered data.  The black line is derived
  from the raw (signal plus noise) data, whereas the red line is the estimate
  of the signal-subtracted PSD. The scan frequency for this chunk appears at
  12 mHz, and the \cmb dipole (which appears as a triangle wave at the scan
  frequency) has been subtracted from the TOD prior to the noise estimation.
  \emph{Right panel:} the amplitude of the noise bias as determined from an
  ensemble of signal plus noise simulations.  The blue line is representative
  of the bias in a typical high signal-to-noise chunk, whereas the red line is
  the most extreme example found in the low signal-to-noise regime. Further
  discussion of the noise in bolometric detectors can be found in Appendix
  \ref{sec:bolonoise}.}
\label{fig:noise}
\end{center}
\end{figure}

To correct for the bias present in the \btk noise estimates, we generate an
ensemble of signal and noise simulations using a 
fiducial noise power spectrum and run the noise estimation procedure
on each realization. The transfer function of the noise estimation 
procedure is then obtained by comparing the ensemble average of the 
estimated noise power spectra to the input power spectra.  The size of 
the ensemble is determined by the required reduction of the sample
variance at the lowest frequencies of interest; we find that the transfer
function is characterized at the sub-percent level with seventy-five
realizations. This bias transfer function is then used to correct the
spectra obtained for each chunk of the time ordered data.
A comparison of bias transfer functions that are typical of data in the high
and low signal to noise regimes is shown in Figure \ref{fig:noise}.

In order to produce noise realizations which accurately reflect the
statistical properties of the instrumental noise, we require a framework
in which to treat noise correlations \emph{between} 
detectors in the time domain.   Noise correlations in the data
are expected both from fundamental considerations (see Appendix
\ref{sec:bolonoise}), as well as from the presence of correlated
thermal/optical fluctuations, and crosstalk in the readout electronics.  
The measured noise from a given channel, $\widetilde{n}^k$, is modeled as the
sum of an intrinsic (uncorrelated) component, $n^k$, and the contributions
from the intrinsic noise of the other channels, filtered through a (frequency
dependent) crosstalk transfer function $\xi_{ik}$.
$$\widetilde{n}^k = n^k + \sum_{i \neq k} \xi_{ik} n^i $$
where, by definition, the intrinsic noise at each frequency is
distributed as an uncorrelated Gaussian distribution,
$$\langle n^i n^k\rangle \equiv \delta_{ik} P_{ik}. $$
The observable quantities
$$\langle \widetilde{n}^i \widetilde{n}^k \rangle = \widetilde{P}_{ik}$$
are the $(N_{ch}(N_{ch}-1)/2 + N_{ch})$ auto- and cross-correlations of the
signal-subtracted time streams, which are estimated directly from the time
ordered data. After correcting for bias, the $\widetilde{P}_{ik}$ are used to
generate realizations of the noise time streams which exhibit the same
correlation structure observed in the data.
These noise realizations are constructed bin-by-bin in the Fourier domain.
For each discrete frequency, we calculate the Cholesky
factorization, $\mathbf{\widetilde{H}}(f)$, of the complex (Hermitian positive
definite) channel correlation matrix,
$$\mathbf{\widetilde{P}}(f) = \mathbf{\widetilde{H}}(f)\mathbf{\widetilde{H}}^T(f).$$
Independent realizations of white noise are generated for each channel.
Simulated data with the proper correlation structure are obtained by operating
on the transform of these realizations with the $N_{ch}\times N_{ch}$ matrix
$\widetilde{H}_{ik}(f)$ for each frequency bin in a given noise-stationary
subset of the data.  Once all of the frequency components are calculated, the
inverse transform provides a correlated noise time stream for each channel
that is used in the Monte Carlo pipeline.


\subsection{Polarized mapmaking}\label{sec:rawsig}

Estimates of the $I$, $Q$, and $U$ parameters can be recovered by
generating orthogonal linear combinations of the data.  For each
sample, $i$, of a given detector and a measurement of the
projection of the orientation of that detector on the sky, $\psi_i$,
one can construct the polarization decorrelation matrix defined by, 
\begin{equation}
\left(\begin{array}{c} d_i \\ d_i \gamma c_i \\ d_i \gamma s_i
\end{array}\right)
= \left(\begin{array}{ccc}
1 & \gamma c_i & \gamma s_i \\
\gamma c_i & \gamma^2 c_i^2 & \gamma^2 s_i c_i \\
\gamma s_i & \gamma^2 c_i s_i & \gamma^2 s_i^2
\end{array}\right)
\left(\begin{array}{c} I \\ Q \\ U
\end{array}\right)~,\label{eqn:decor}
\end{equation}
where $\gamma\equiv \frac{(1-\epsilon)}{(1+\epsilon)}$ is a
parameterization of the polarization efficiency.
For simplicity we have abbreviated the trigonometric
functions, whose argument is $2\psi_i$.

In the limit that the instrumental noise time stream, $\mathbf{n}$, is
stationary, Gaussian, and is well-characterized by a white frequency
spectrum, the optimal map is obtained by summing all time  
samples $d_i$ and decorrelation matrix elements falling in a pixel
$p$.  Assuming that the scan strategy, instrument, or channel
combination provides modulation of the angle $\psi$, the matrix is
nonsingular and the best estimates for $I,Q$, and $U$ are then
obtained by inverting the coadded (3x3) decorrelation matrix at each
pixel. This is the polarized analog to a naively coadded temperature map.

The situation becomes markedly more difficult in the presence of
noise with nontrivial statistics.  The solution that is optimal in the least
squares sense is again given by Equation \ref{eqn:map}, with the
understanding that now the data consist of the linear combinations
defined by Equation \ref{eqn:decor}.
We now turn to the problem of finding the solution to
Equation \ref{eqn:map} for polarized data, in the presence of noise with
unknown statistical properties, using channels of varying
sensitivity and polarization efficiency. In this general case, the
data are treated in the following way:  estimates of the left hand
side of the \emph{noise only} version Equation \ref{eqn:decor} and 
the Stokes decorrelation matrix are generated for each pixel,
\begin{equation}
\hat{n}_{p} = \sum_j^{N_{ch}} w_j ~\sum_{i\in p} 
\left(\begin{array}{c} n_i \\ n_i ~\gamma_j ~c_i \\ n_i ~\gamma_j ~s_i
\end{array}\right)~,
\label{eqn:ntod}
\end{equation}
where the $n_i$ are the elements of signal subtracted time stream. Likewise,
for the decorrelation matrix one calculates 
\begin{equation}
\hat{M}_{p} = \sum_j^{N_{ch}} w_j ~\sum_{i\in p} 
\left(\begin{array}{ccc} 1 & \gamma_j c_i & \gamma_j s_i \\
            - & \gamma_j^2 c_i^2 & \gamma_j^2 c_i s_i \\
            - & - & \gamma_j^2 s_i^2 \\
\end{array}\right)~.
\label{eqn:mmatrix}
\end{equation}

One then obtains an estimate of the {\em corrections} to the Stokes
parameters $I$, $Q$, and $U$ maps for an iteration $k$, by
inverting $\hat{M}$ at each pixel,
\begin{equation}
\mathbf{S}_{k+1}-\mathbf{S}_k = \mathbf{\hat{M}}_k^{-1}
\mathbf{\hat{n}}_k~,
\label{eqn:pmap}
\end{equation}
allowing one to iteratively obtain a solution for the maximum
likelihood maps of each Stokes parameter. 

\begin{figure}[!t]
\begin{center}
\rotatebox{90}{\scalebox{0.35}{\includegraphics{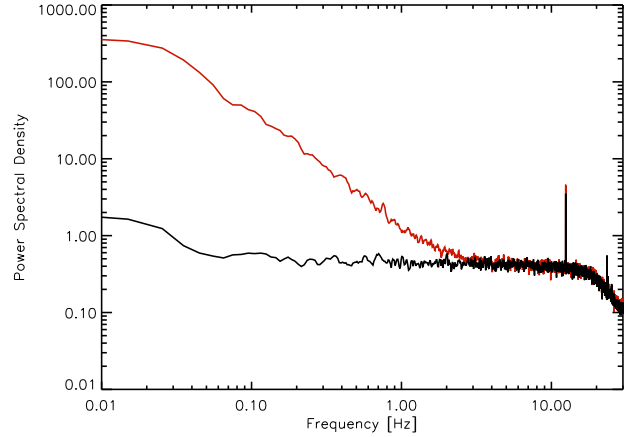}}}
\caption[PSB common mode rejection]{\small The power spectrum of 
  sum and difference time streams from a typical 145 GHz PSB pair observing
  (apparently) unpolarized atmospheric fluctuations during the austral summer
  in Antarctica.  When measuring a small polarized signal buried in a large
  unpolarized background, the high degree of common mode rejection of the PSBs
  makes them naturally suited to an analysis of the sum and difference
  time streams, as described in Section \ref{sec:sumdiff}.}
\label{fig:cmrr}
\end{center}
\end{figure}

\subsection{Sum and difference time streams}\label{sec:sumdiff}

An alternate approach to signal and noise estimation involves operations on
the sum and difference of the calibrated time streams from bolometers within a
PSB pair. This has the numerical advantage of isolating the temperature and
polarization terms in the numerical inversion of Equation
\ref{eqn:mmatrix}. This approach takes full advantage of the high degree of
common mode rejection of the PSB design, which is illustrated in Figure
\ref{fig:cmrr}. The advantages of this approach, which will be discussed in
more detail in Section \ref{sec:resid}, are obtained at the cost of suboptimal
noise weighting of the channels within a pair.

We may represent a sample, $i$, of a single detector as the linear combination
of the sort,
\begin{equation}
s_i = I + \gamma_i \left(Q\cos 2\psi_i + U\sin 2\psi_i\right)~.
\end{equation}
Assuming that the channels are properly calibrated, the sum and
difference of the signals from a PSB pair may be written as,
\begin{equation}
{}^+s_i\equiv\frac{1}{2}(s_1+s_2)_i= I +\frac{1}{2}({}^+\alpha_iQ+{}^+\beta_iU)
\end{equation}
\begin{equation}
{}^-s_i\equiv\frac{1}{2}(s_1-s_2)_i=\frac{1}{2}({}^-\alpha_iQ+{}^-\beta_iU)~,
\end{equation}
where we have defined the angular coefficients
\begin{eqnarray}
{}^\pm\alpha_i &=& \gamma_1\cos 2\psi_{1i}\pm\gamma_2\cos 2\psi_{2i} \\
{}^\pm\beta_i &=& \gamma_1\sin 2\psi_{1i}\pm\gamma_2\sin 2\psi_{2i}
\end{eqnarray}
in terms of the independent variables, $\psi_{ki}$, where $k=\{1,2\}$
identifies the channel. Recall that for a PSB pair
the angular separation of the channels is $\Delta \simeq 90 \pm
2^\circ$, however this treatment in no way requires that to be the case.

Following the prescription of Section \ref{sec:sig}, one generates linear
combinations of the differenced data,
\begin{equation}
\left(\begin{array}{c} {}^-s_i {}^-\alpha_i \\ {}^-s_i {}^-\beta_i
\end{array}\right) = \frac{1}{2}\left(\begin{array}{cc} 
{}^-\alpha_i^2 & {}^-\alpha_i{}^-\beta_i \\
{}^-\alpha_i{}^-\beta_i & {}^-\beta_i^2 \end{array}\right)
\left(\begin{array}{c} Q \\ U \end{array}\right)
\end{equation}
As before, one builds up information about the $Q$, $U$ decorrelation
matrix through the combination of channel pairs, as well as modulation
of the angular coverage, $\psi$.  In this regard we have
\begin{equation}
{}^-\hat{n}_{p} = \sum_j^{N_{pairs}} w_j ~\sum_{i\in p} 
\left(\begin{array}{c} {}^-n_i{}^-\alpha_i\\ {}^-n_i {}^-\beta_i
\end{array}\right)~,
\label{eqn:dntod}
\end{equation}
where the time-streams ${}^-n_i$ represent the polarization subtracted
difference data.  The $2\times2$ decorrelation matrix is, therefore, 
\begin{equation}
{}^-\hat{M}_{p} = \frac{1}{2}\sum_j^{N_{pairs}} w_j ~\sum_{i\in p} 
\left(\begin{array}{cc}
{}^-\alpha_i^2 & {}^-\alpha_i{}^-\beta_i \\
{}^-\alpha_i{}^-\beta_i & {}^-\beta_i^2  \\
\end{array}\right)
\label{eqn:dmmatrix}
\end{equation}
and we note that we are now using suboptimal weighting of the
\emph{pairs} to generate corrections to the polarization map.  Note
that, for $\Delta \simeq 90^\circ$, the quantities ${}^-\alpha$ 
 and ${}^-\beta$ have opposite parity, so that when averaged over a large
 sampling of $\psi$, the off-diagonals of Equation \ref{eqn:dmmatrix} are
 small. Once the corrections to $Q$ and $U$ are obtained, one may substitute
 them in the sum for ${}^+s$ to solve self consistently for $I$.


\subsection{Polarized cross-linking}\label{sec:resid}

The iterative map-making methods described in Sections \ref{sec:rawsig} and
\ref{sec:sumdiff} result in a self-consistent estimate of the signal and noise
from the data that is ``optimal'' in the least-squares sense.  However,
instrumental effects can introduce correlations in the \emph{signal}
contribution to the time domain data that limit the fidelity of the recovered
Stokes parameter maps. \footnote{Common
  examples of such instrumental effects include the impact of the AC coupling
  of detector outputs, variations in the noise spectra between detectors, scan
  synchronous noise, and limited accuracy of estimates of the low frequency
  noise. These noise estimates are fundamentally sample variance limited by the
  finite period over which the noise can be considered to be stationary.} 

In simulations using noise correlations to process the \emph{signal only}
time streams according to Equation \ref{eqn:map}, these effects appear as a
residual between the input and recovered Stokes parameter maps.
The spatial morphology and
amplitude of these residuals depend on the amount of cross-linking in the scan
strategy, the degree of polarization modulation, as well as the method used to
decorrelate the $I$, $Q$, and $U$ parameters from the time stream.
While these residuals do not introduce a bias to the
pseudo-$\mathcal{C}_\ell$ estimates of the power spectra, they do contribute
to the signal covariance of the map, and therefore degrade the sensitivity of
the Monte Carlo approach relative to optimal methods.\footnote{ The residuals
  contribute directly to the effective transfer function for the temperature
  and polarization spectra (the $\mathcal{F}_\ell$ discussed in \citet{master}
  and \citet{contaldi}).} 

The fidelity of the recovered Stokes parameter maps is an important
consideration for the design of scanning polarimeters; the statistical depth
of the survey determines the level at which these instrumental artifacts must
be controlled.
To investigate these effects, we generate signal-only simulations based
on the \btk observation strategy and the measured \btk noise power
spectra.  The \btk \cmb data consist of a deep region and a shallow region,
representing the extreme cases of possible observation strategies avaliable to
Antarctic LDB payloads. The \btk scan crosses each pixel in the
deep survey over may timescales and at many different orientations (due to sky
rotation), while the pixels in the shallow survey are not well sampled.

\begin{figure}[!t]
\begin{center}
\rotatebox{0}{\scalebox{0.35}{\includegraphics{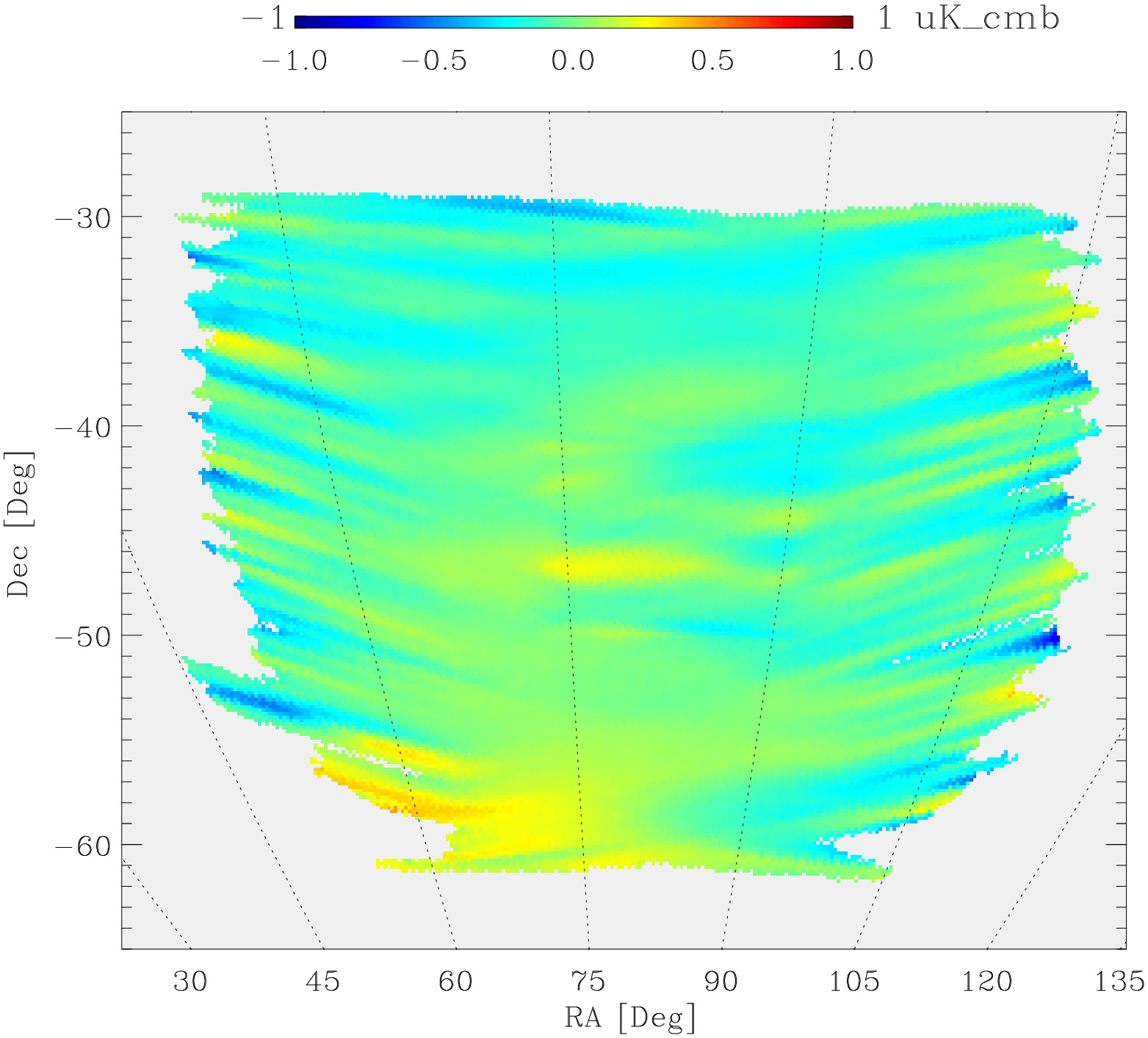}}}
\rotatebox{0}{\scalebox{0.35}{\includegraphics{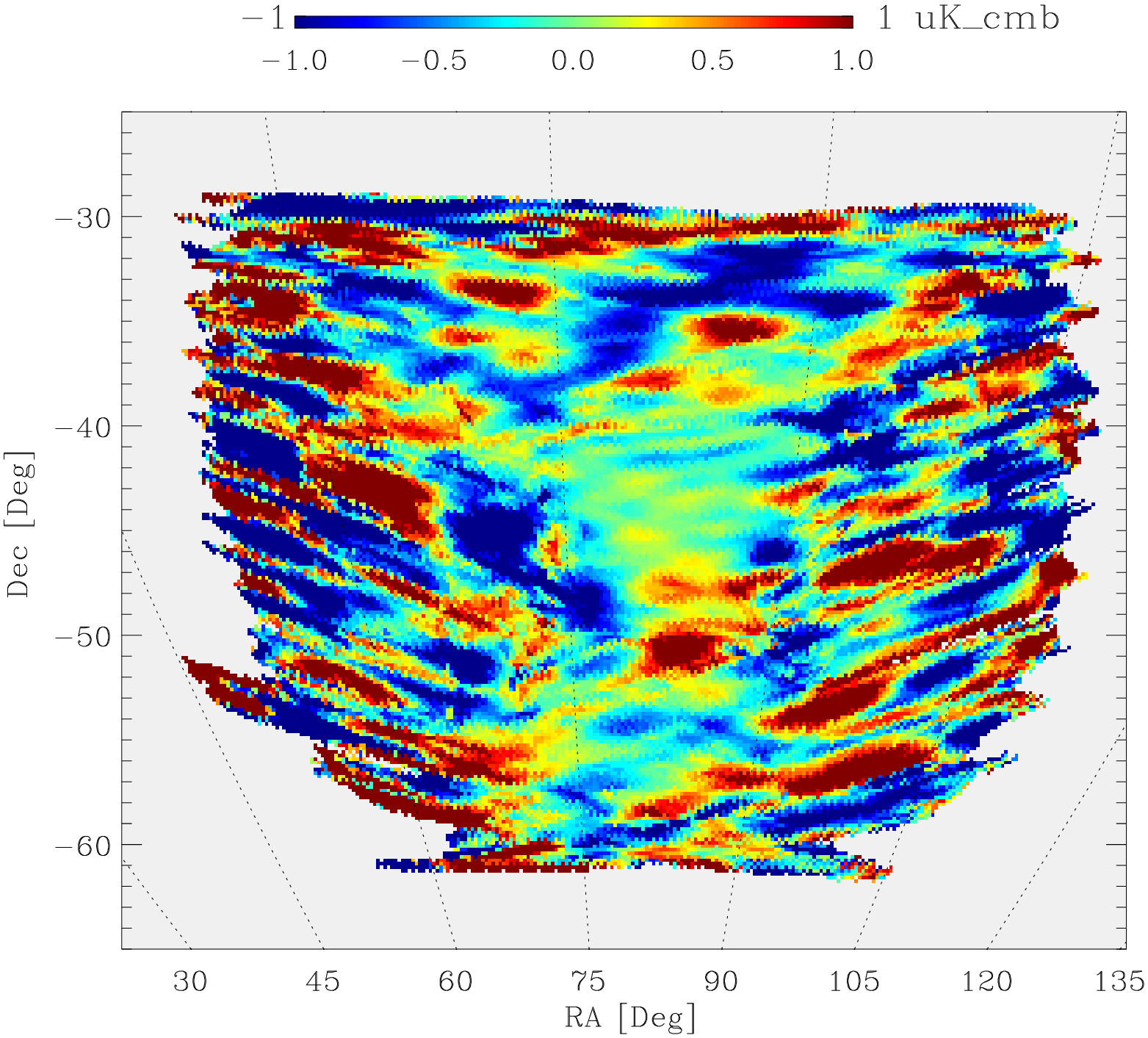}}}
\caption[Comparison of direct and differencing methods]{\small A
  signal-only simulation, showing the residuals between the observed
  and input polarization (in this case, Stokes $Q$). The
  differencing method of Section \ref{sec:sumdiff} \emph{(left panel)} is
  more robust to common mode effects than is the more general method of Section
  \ref{sec:rawsig} \emph{(right panel)}, especially in regions where the
  crosslinking is poor. This is due
  primarily to the correlations that are introduced by the preconditioning of
  the TODs (essentially a highpass filter at 20 mHz) and the features in the
  noise kernel $N^{-1}$, which introduce path dependencies to the observed
  $I$, $Q$, and $U$ parameters.} 
\label{fig:jiqudiqu}
\end{center}
\end{figure}

Using the Healpix \emph{synfast} facility~\cite{healpix}, we generate a
noise-free polarized \cmb sky, pixelized at $3.4^\prime$ resolution, from a
concordance  $\Lambda$CDM model.  We then simulate three
polarization modulation schemes to compare with the nominal \btk modulation
(i.e. sky rotation alone). Each time-domain simulation includes the nominal
sky rotation in addition to that which would be achieved with a rotating
half-wave plate. We model the following modulation schemes, which are
representative of those proposed by balloon borne and terrestrial bolometric
polarimeters~\cite{ebex},

\begin{enumerate}
\item 22.5$^\circ$ steps of the polarization angle each hour.
\item 22.5$^\circ$ steps of the polarization angle at the end of every scan. 
\item Continuous rotation of the polarization angle of each PSB at 350
  mHz.\footnote{We choose this modulation rate to be as fast as possible,
  given the \boom scan and sample rates.} 
\end{enumerate}

We observe the simulated sky with each of these polarization
modulation schemes and create a noise-free time ordered data set, $s$, for
each. We then solve for the signal part of the general least squares map using
the \btk inverse noise filters, $N^{-1}$, according to Equation \ref{eqn:map}:
$$
\tilde{m} = \left( A^T N^{-1} A \right)^{-1} A^T N^{-1} s
$$
The noise kernels, $N^{-1}$, are smoothly truncated below 70 mHz.\footnote{The
  lowest frequency that can be reliably recovered clearly affects the
  amplitude of the residuals, especially for the temperature fluctuations
  which have significant power on large scales.  However, the relative
  benefits of the scanning strategies outlined above are generally insensitive
  to the exact value of the minimum frequency.} We decorrelate the Stokes
$I$,~$Q$, and $U$ parameters at $6.8^\prime$ resolution, using both the
general ($3\times 3$) method and the PSB sum/difference ($2\times 2$) method,
and compare the resulting polarization maps with the input sky.  In the case
of the former method, the residuals in the $Q$ and $U$ maps contain
contributions from the finite resolution of the pixelization as well as the
correlations introduced in the course of making maps from the time-ordered
data. The difference time streams do not contain the relatively large
unpolarized contribution, and therefore are far less susceptible to these
pixelization effects.

\begin{figure}[!p]
\begin{center}
\rotatebox{0}{\scalebox{0.22}{\includegraphics{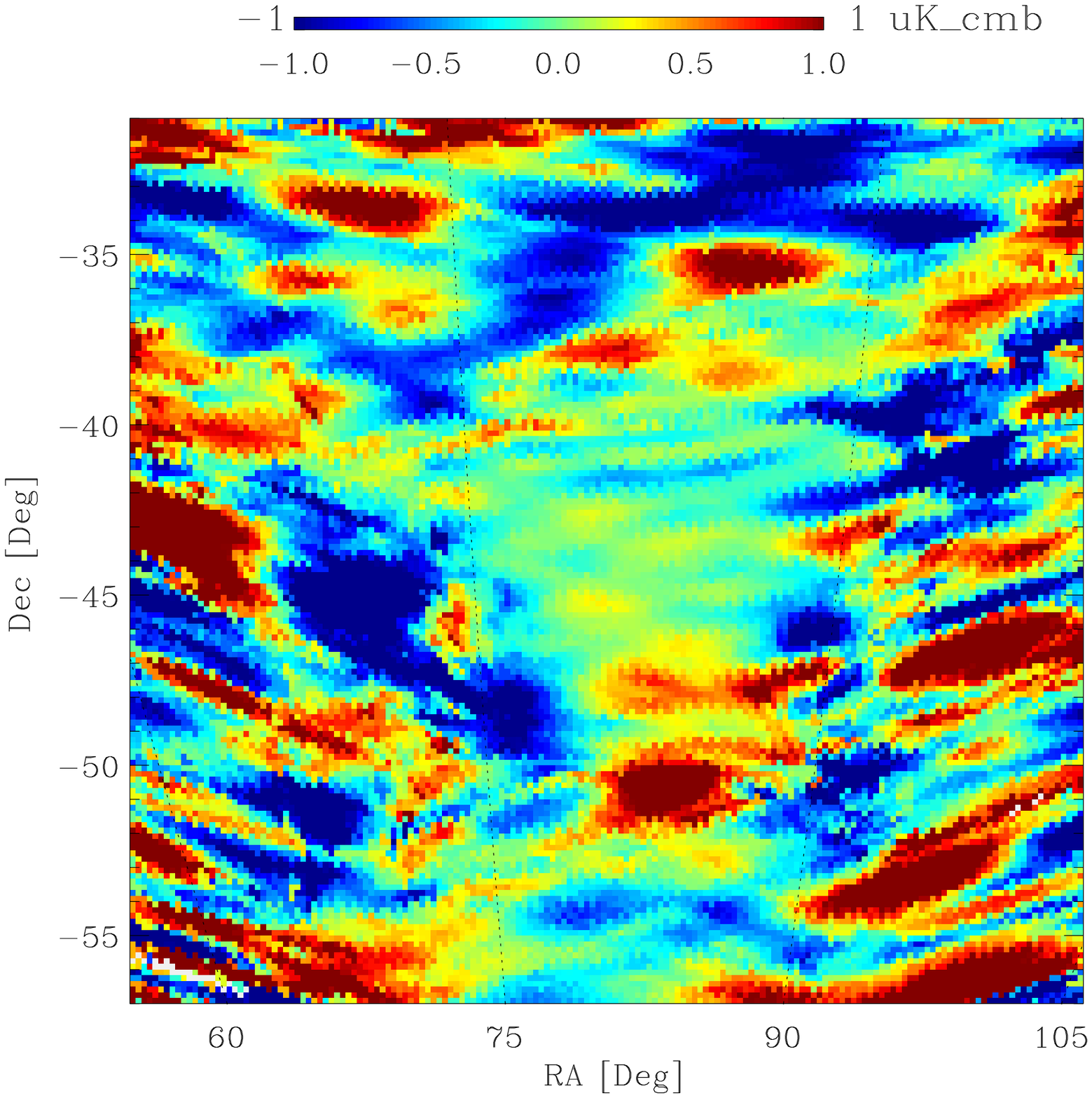}}}
\rotatebox{0}{\scalebox{0.22}{\includegraphics{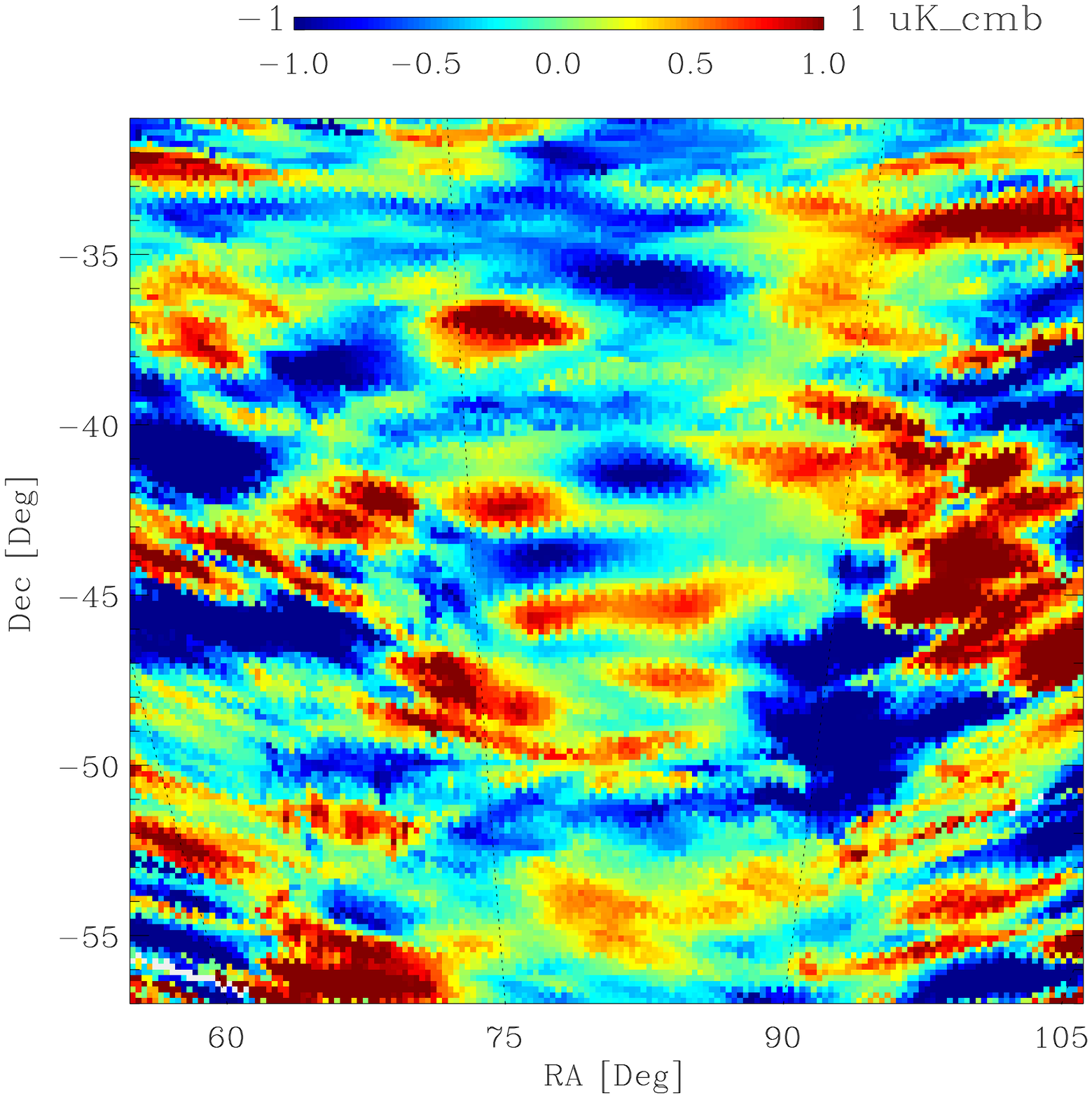}}} \\
\rotatebox{0}{\scalebox{0.22}{\includegraphics{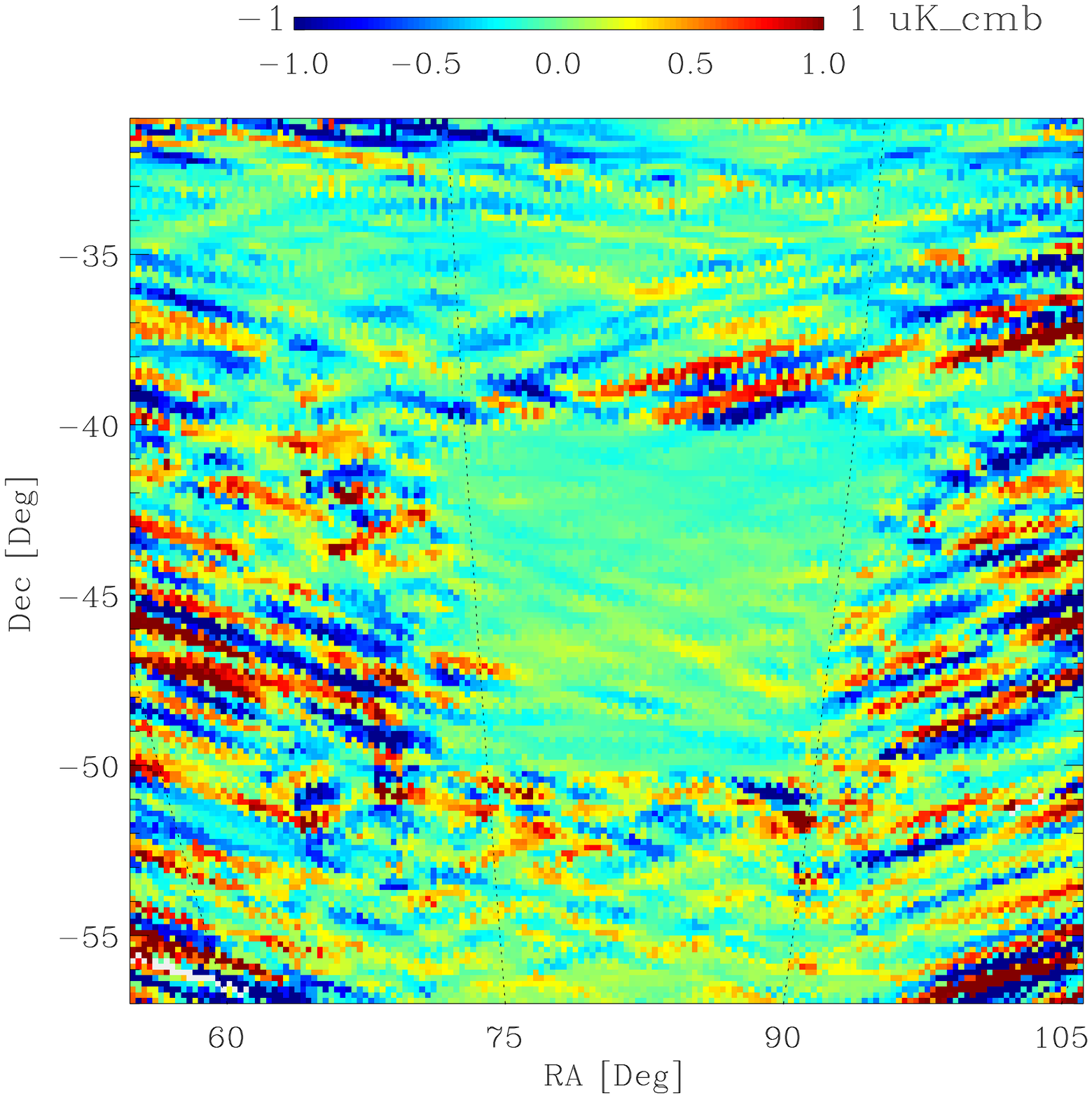}}}
\rotatebox{0}{\scalebox{0.22}{\includegraphics{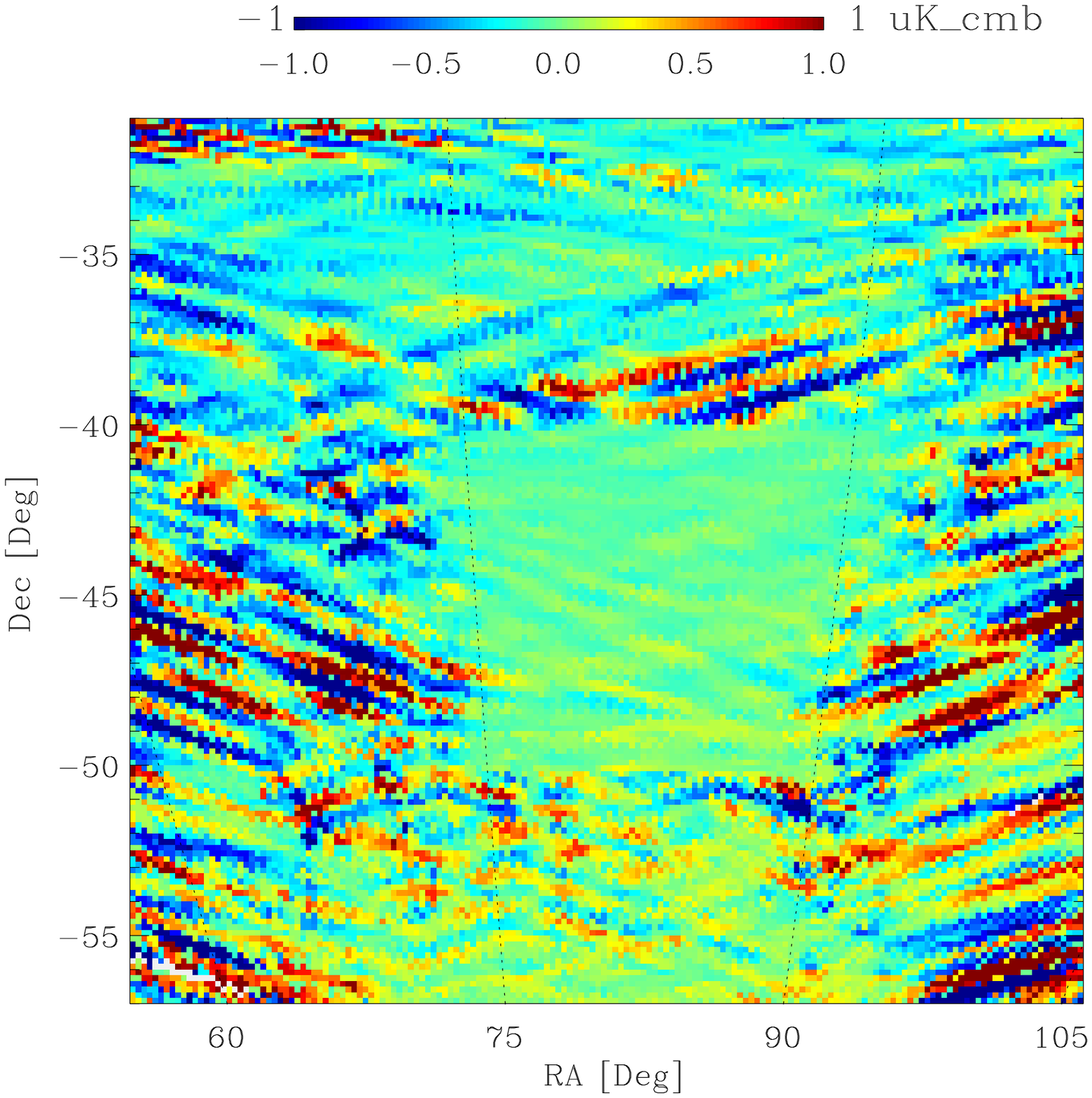}}} \\
\rotatebox{0}{\scalebox{0.22}{\includegraphics{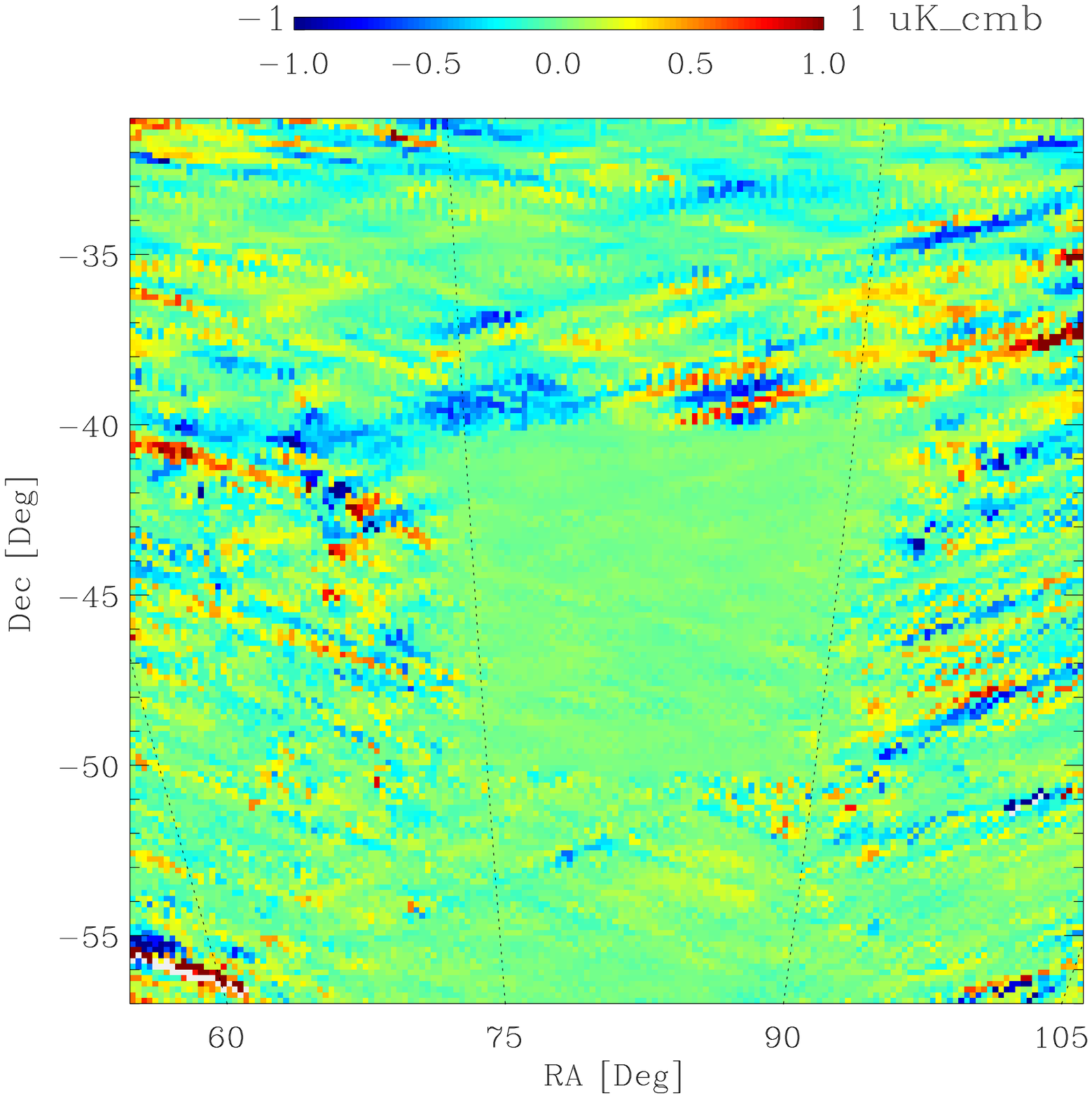}}}
\rotatebox{0}{\scalebox{0.22}{\includegraphics{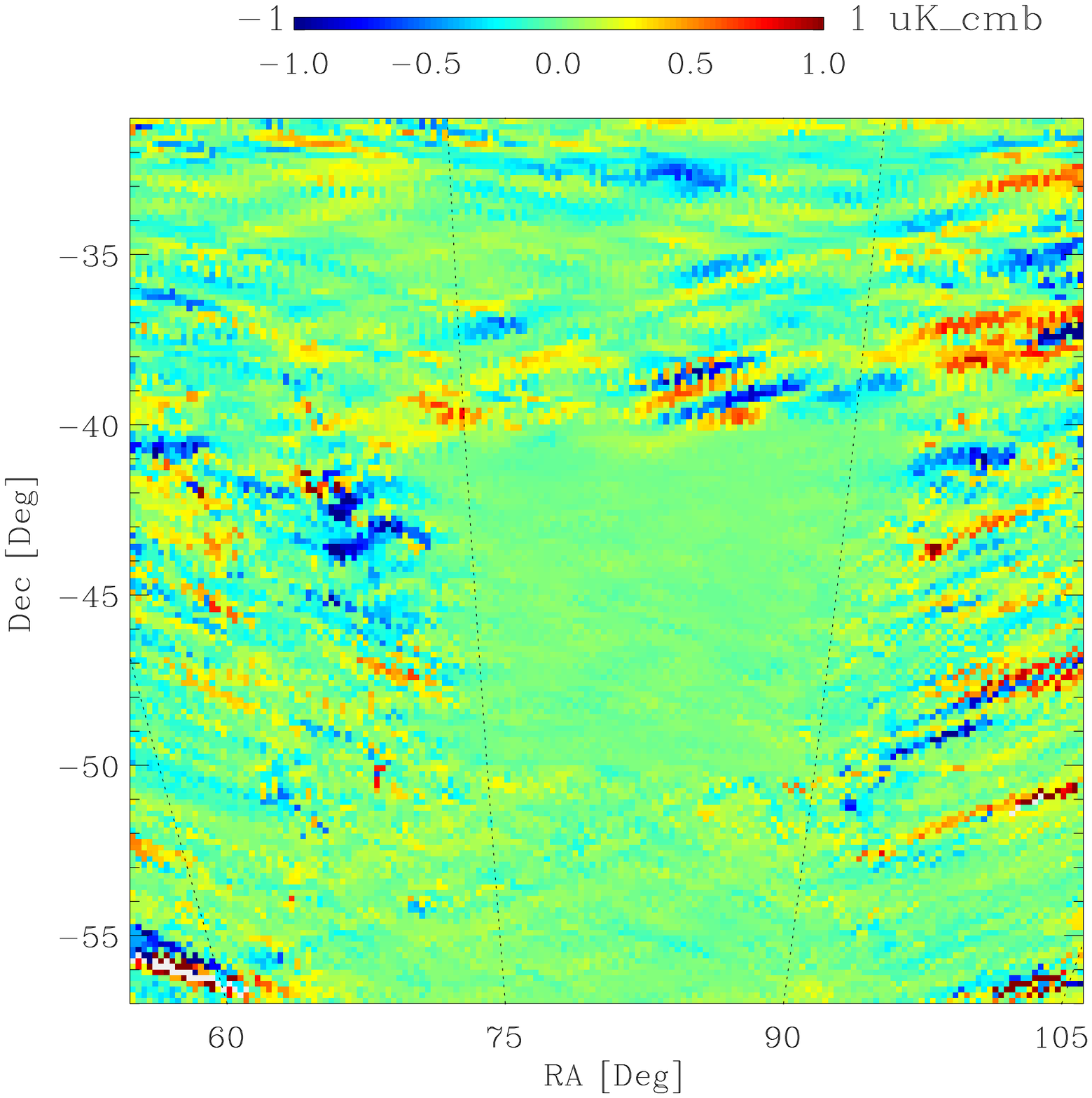}}} \\
\rotatebox{0}{\scalebox{0.22}{\includegraphics{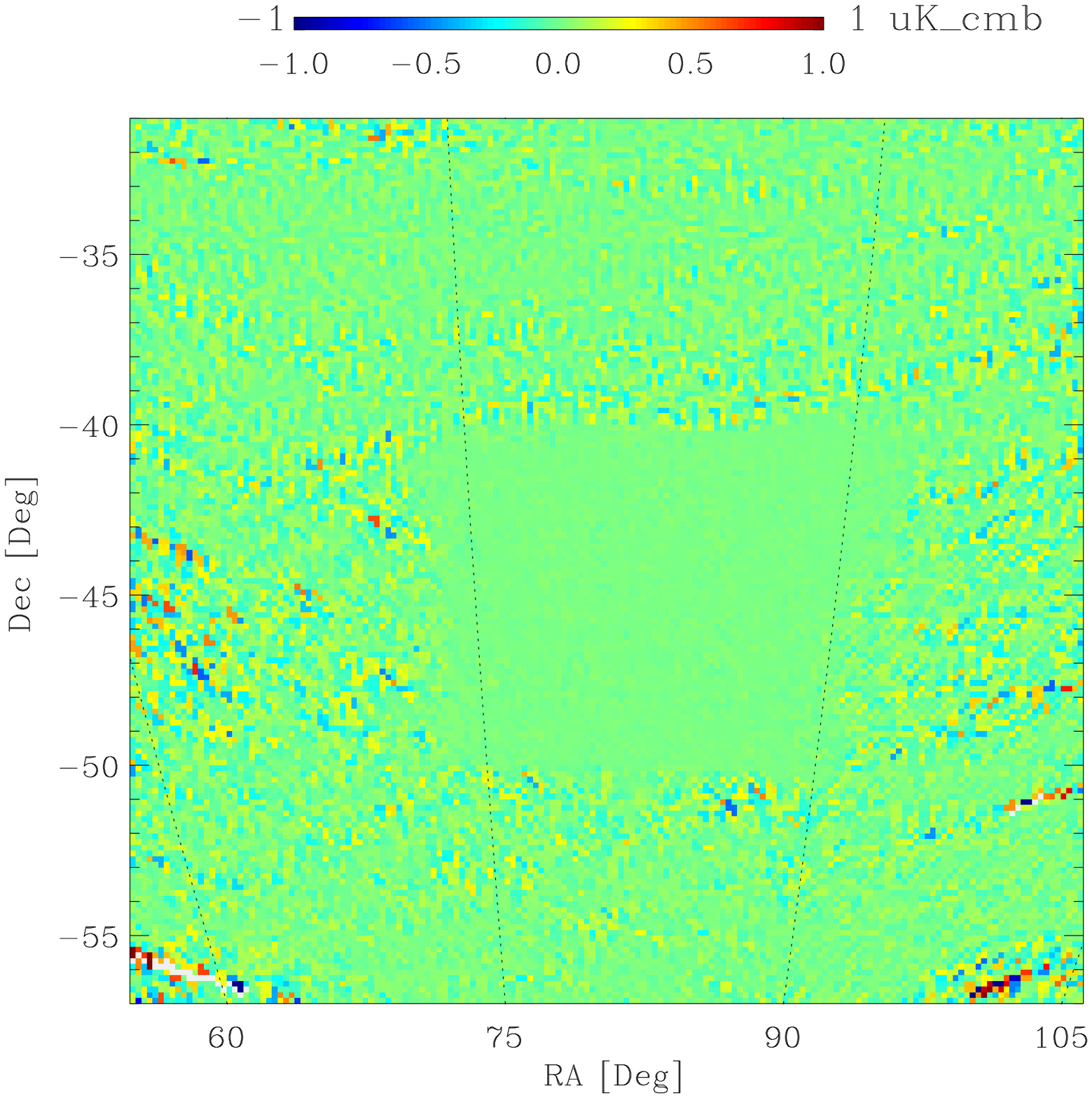}}}
\rotatebox{0}{\scalebox{0.22}{\includegraphics{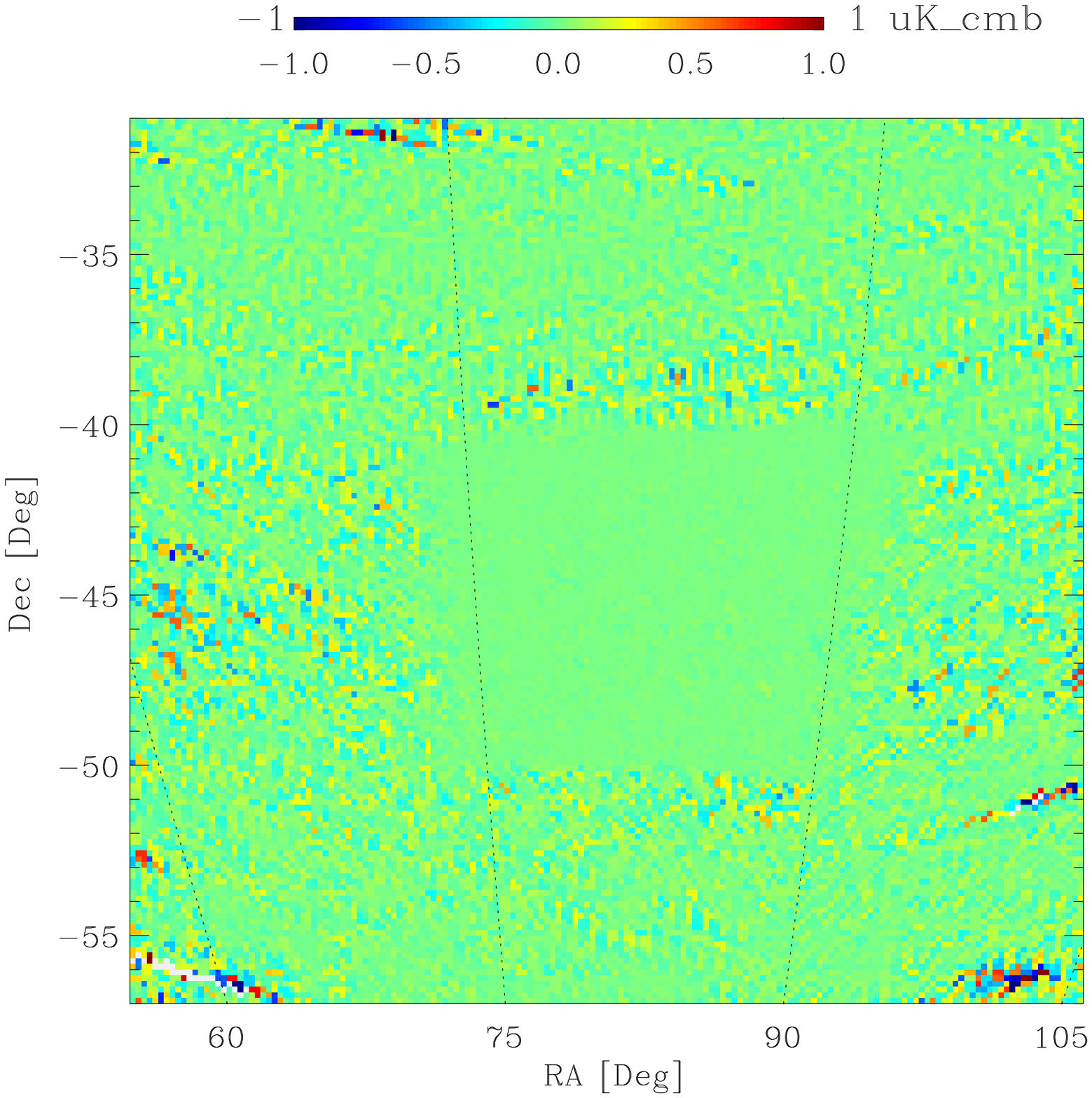}}}
\end{center}
\caption{The Joy of Crosslinking.  The residual signal in the Stokes $Q$
  \emph{(left column)} and $U$ \emph{(right column)} parameter maps generated
  using the general algorithm of Section \ref{sec:rawsig}, for increasing
  levels of polarization modulation. \emph{At top}, the only modulation is
  that provided by sky rotation.  The \emph{middle} two sets are the residuals
  obtained when stepping the half waveplate by $22.5^\circ$ ($Q\rightarrow U$)
  each hour and at the end of each azimuth scan, respectively.  The
  \emph{bottom} row shows the fidelity achieved with a waveplate spinning
  continuously at 350 mHz. The remaining residuals are dominated by
  pixelization effects.}  \label{fig:psimodsim_maps}
\end{figure}

Before considering the effects of the polarization modulation, we first
investigate the benefits of exploiting the common mode rejection of the PSB
pairs through the analysis of the difference time streams.  We compare the
residuals resulting from the application of the general method of
Section \ref{sec:rawsig} and from that of Section \ref{sec:sumdiff}. 
In Figure \ref{fig:jiqudiqu} we show the qualitative improvement in the
fidelity of the reconstruction that results from the analysis of the
difference time streams.  It should be noted that, even for the general method,
the sky rotation of the nominal \btk scan provides a degree of
modulation that is sufficient to reduce the residuals to a level well below
that of the instrumental noise in the \btk maps ~\cite{b2k_tt,b2k_inst}; the
use of a waveplate in \btk would not have significantly improved the accuracy
of the polarimetry. 
Furthermore, the direct difference method of Section
\ref{sec:sumdiff} is less sensitive to the limited cross-linking of the
nominal scan than is the general polarization decorrelation method of Section
\ref{sec:rawsig}.  

While the design of the PSBs is naturally suited to the sum/difference
approach, scanning experiments sensitive to a single polarization (such as
\ebex\ \cite{ebex} and \spider\ \cite{spider}) are not able to exploit the
common mode rejection that is intrinsic to the design of the PSBs.  A scheme
for polarization modulation is therefore a highly desirable feature in singly
polarized systems.

In order to illustrate the effect of the polarized cross-linking on the
fidelity of the reconstructed signal, we show in Figure
\ref{fig:psimodsim_maps} the residuals that result, for
a particular \cmb realization, from the general ($3\times 3$) 
approach of Section \ref{sec:rawsig} for the nominal \btk scan, and for each
of the three modulation schemes listed above.  The fidelity of the signal
reconstruction improves in proportion to the rate of modulation.  The improved
cross-linking randomizes the path dependencies of the observed Stokes
parameters that are introduced by the processing of the time ordered data,
namely Equation \ref{eqn:map}.

\begin{figure}[!t]
\begin{center}
\includegraphics[angle=90,width=12cm]{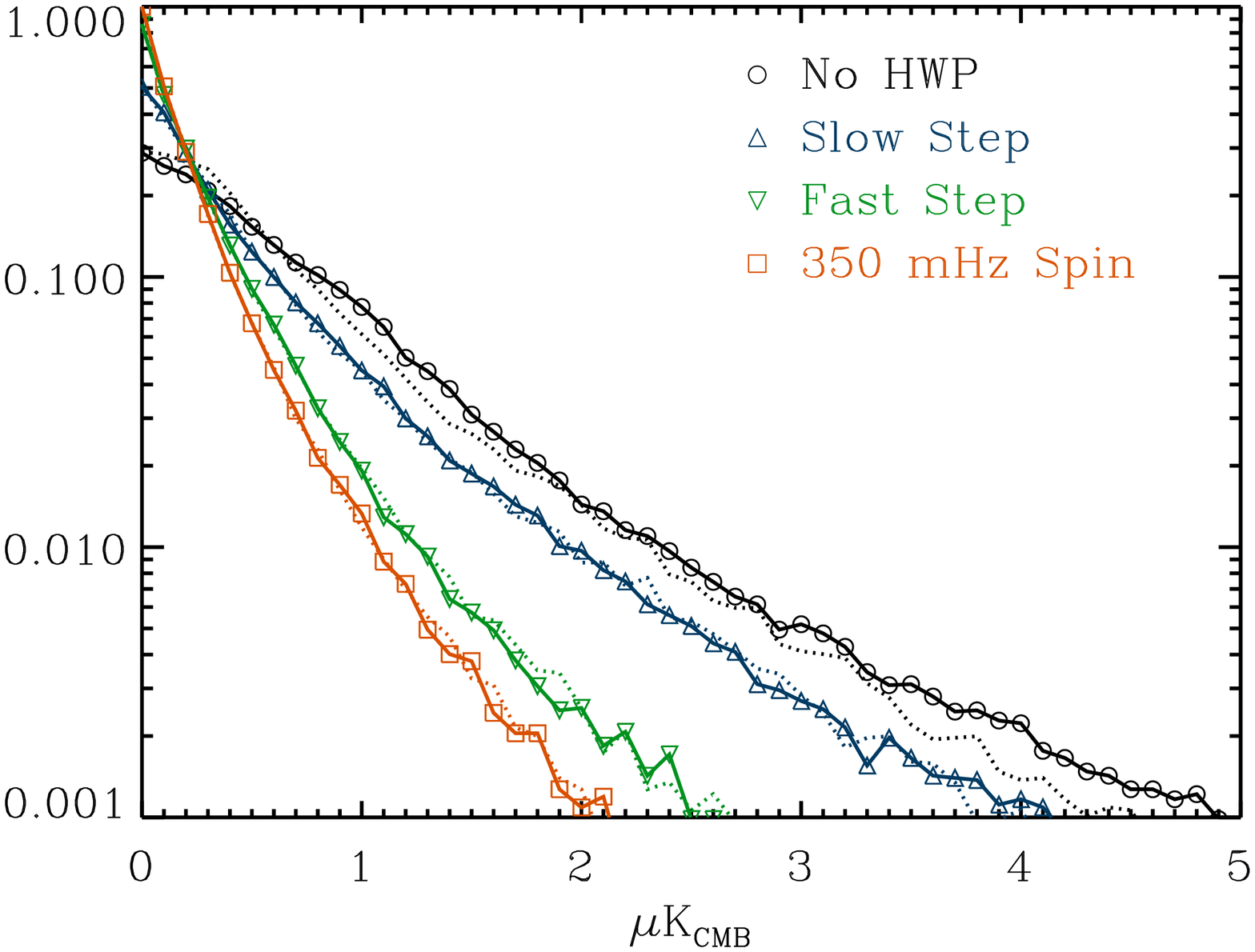}
\end{center}
\caption{\small 
Histograms of the Stokes $Q$/$U$ residuals, using $6.8^\prime$ pixels, for
various modulation schemes.  Figure \ref{fig:psimodsim_maps} shows that the
errors are largest on large scales.  The residuals in the last row are
dominated by pixelization effects.}
\label{fig:quhist}
\end{figure}

As shown quantitatively in Figure \ref{fig:quhist}, the largest residuals in
the $Q$ and $U$ maps (which, in the B03 example, occur on relatively large
scales) can be significantly reduced by a modest degree of polarization
modulation. Nevertheless, for the effects that have been included in this
simulation, generating maps using the sum and difference time streams of the
PSBs is nearly as effective at minimizing the residuals in the $Q$ and $U$
maps as the use of a half wave plate. It is important to note that beam
asymmetries, instrumental polarization, pointing errors, and calibration
uncertainties are examples of effects that are \emph{not} included in these
simulations. Each of these effects are mitigated by the use of an idealized
modulation scheme, but not by the sum/difference method of Section
\ref{sec:sumdiff}.


\section{Summary and Conclusions}\label{sec:con}

We have described in detail the design and performance of the Polarization
Sensitive Bolometers (PSBs) which have enabled the first generation of
successful bolometric \cmb polarimeters.  This discussion outlines the
instrument parameters which must be characterized to accurately decorrelate
the Stokes $I$, $Q$ and $U$ parameters from the time ordered data of a PSB.
The design of the PSBs provides a high degree of common mode rejection that
can be exploited in the analysis to minimize susceptibility to various
instrumental effects that can potentially limit the fidelity of the recovered
polarization. Simulations of PSB data including realistic instrumental effects
illustrate the benefits of analyzing difference time streams, as well as
various practical schemes for modulating the polarization signal.


\begin{acknowledgements}
The authors would like to thank Eric Hivon for many helpful discussions.  The
bolometers used by {\boomn},{\quest},{\bicep}, and {\planckhfin} were
fabricated at the Micro--Devices Laboratory at JPL (Jamie Bock and Anthony Turner). {\boom} is an international collaboration
between Caltech, JPL, IPAC, LBNL, U.C. Berkeley, Case Western Reserve
University, University of Toronto, CITA, Cardiff University, Imperial College,
Universit\`a di Roma, INGV, INFN, and Institut d\rq Astrophysique.

\end{acknowledgements}

\clearpage
\bibliographystyle{aa}

\newpage
\appendix

\section{Noise in bolometric receivers}\label{sec:bolonoise}

\subsection*{General noise properties}


Noise in bolometric receivers originates from several independent
sources, including contributions from the readout electronics, the detector,
and the intrinsic fluctuations in the optical background power.  These noise
sources are independent processes and their contributions add in quadrature to
the total noise of the system.

The contribution of each of these components to the time ordered data are
filtered by the transfer function of the one or both of the detector and the
readout electronics\footnote{For contemporary bolometric instruments like
  \planckhfin, the readout electronics include a cold JFET amplifier, ambient
  temperature amplifier/bandpass filters, and an anti-aliasing/data
  acquisition system.  In practice, the contribution to the noise of
  everything except the JFET amplifiers and low noise preamplifier are
  negligible.}.  The voltage noise (that is, the Johnson, JFET/amplifier
noise, and the product of current noise with the series impedance) of the
system, $n_v$, is filtered only by the 
transfer function of the readout, $\widetilde{Z}(f)$. The photon noise,
$n_\gamma$, and phonon noise, $n_G$, are filtered not only by the bolometer
voltage responsivity, $\widetilde{S}(f)$, but also by the readout.  
The bolometer transfer function, $\widetilde{S}$, is typically that of a
single low pass filter, or a cascade of two such filters.  The details of the
readout transfer function, $\widetilde{Z}$, vary, but always includes an
anti-aliasing filter which strongly attenuates frequencies well below the
Nyquist frequency of the analog-to-digital converter (ADC). \footnote{In some
  instances, the electronics are AC coupled, meaning that the low frequencies
  are strongly attenuated. The advent of low-cost, high-resolution ADCs has
  made this feature less common. }
The raw data are composed of the signal from the sky, $s$, and the
various noise contributions, convolved with the bolometer and readout transfer
functions,
\begin{equation}
d = Z~\otimes~\left[ ~S~\otimes 
~\left( s+n_\gamma+n_G \right) + n_v \right] .
\end{equation}
One of the first stages of analysis involves the deconvolution, and then
the de-glitching, of this raw detector time stream.  The deconvolution is
 normally accomplished in the Fourier domain by dividing the \emph{product}
 of the bolometer and electronics transfer functions, $\widetilde{Z}^\prime =
\widetilde{Z}~\widetilde{S}$, from the raw data, $d$. While this results in a
time stream that is characterized by a signal component with a uniform
calibration in the frequency domain, the contribution of the voltage
noise is biased according to the detector transfer function, 
$\widetilde{n}_v~\rightarrow~\widetilde{n}_v/\widetilde{S}$.
Because the bolometer transfer function has the form of a low-pass
filter, the deconvolved time stream generally exhibits \lq\lq$f$-noise\rq\rq\ in
proportion to the time constant of the bolometer and the amplitude of the
voltage noise component.

Figures \ref{fig:cmrr} and \ref{fig:sonogram} are examples of the power
spectrum of a time stream prior to deconvolving the system transfer function.
Figure \ref{fig:noise} shows the power spectrum of the deconvolved data. A
precise knowledge of the system transfer function is necessary to avoid the
introduction of instrumental artefacts in the recovered signal. In the case of
CMB studies, such an error generally results in a bias in power spectra
derived from the maps~\cite{b2k_tt}.

Contemporary bolometric receivers, even those operating in the low
background environment provided by balloon and orbital payloads, are designed
to achieve background limited sensitivities. In these receivers photon noise
is a major, if not dominant, contribution to the total noise in the system. In
the following section we examine aspects of this photon noise contribution,
$n_\gamma$, including a derivation of the (low level) noise
correlations expected between detectors in a PSB pair resulting from
fundamental properties of fluctuations in the thermal background radiation. 

\subsection*{Photon noise}

A fundamental limitation to the sensitivity of any receiver (band-gap,
coherent, or bolometric) derives from the intrinsic temporal fluctuations in
the optical, often thermal, background radiation.
The noise properties of thermal background radiation, or photon noise,
differ greatly between radio, sub-millimeter, infrared, and optical
instrumentation due to their vastly different operational regimes of
photon occupation number.  Photons satisfy Bose-Einstein statistics,
and therefore the occupation of a mode of frequency $\nu$ is 
\begin{equation}
n(\nu,T)= \frac{1}{e^{h\nu/kT}-1}\label{eqn:bose} 
\end{equation}
for a thermal background of temperature, $T$.

The ratio $k/h= 20.8$ [GHz/K] sets, for a given background
temperature, the frequency for which average occupancy is above
or below unity.  At radio wavelengths astronomical instruments
typically enjoy background levels of order 10 K, with the minimum
background limited by the \cmb monopole at 2.728 K. At higher
frequencies, atmospheric loading and thermal emission from the
instrument tend to dominate the background, and are typically $\sim
30-100$ K for terrestrial telescopes. Therefore,
instruments operating at frequencies above $\sim 100$ GHz have
occupation numbers of order unity, while receivers at lower
frequencies tend to have very large occupation numbers, 
$n\simeq kT/h\nu$. In the low $n$ regime, photons can be thought 
of as arriving at the detector sporadically.  The photon noise in 
high frequency ($\gtrsim 100$ GHz) instruments with low backgrounds 
can therefore be expected to largely satisfy Poisson
statistics, where one expects fluctuations on the mean to scale roughly as
$\sqrt{N}$.

Hanbury Brown and Twiss were the first to complete a rigorous analysis of
noise correlations in photons
\cite{hbt56a,hbt56a,hbt56c,hbt57a,hbt57b}.  The topic has been
continually revisited in the fifty years since the first published
work, and is still relatively un-advertised among many instrumentalists
and observers alike.  Therefore, we go through the analysis in
detail.

\begin{figure}[!t]
\begin{center}
\rotatebox{0}{\scalebox{0.35}{\includegraphics{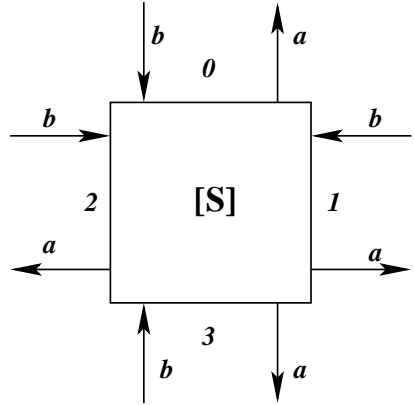}}}
\caption[PSB Scattering matrix]{The scattering matrix of a
  four port network. A polarization sensitive bolometer can be modeled
  by such a network.}
\label{fig:smatrix2}
\end{center}
\end{figure}

Following~\citet{jonas03}, we can write the covariance matrix describing
detector outputs in all generality
\begin{equation}
\sigma_{ij}^2 = \frac{1}{\tau}\int d\nu B_{ij}\left(B_{ji}  +
\delta_{ij}\right)~,
\label{eqn:photon_covar}
\end{equation}
where we define the power coupling matrix
\begin{equation}
B_{ij} \equiv h\nu \sum_k S_{ik}S_{jk}^* n_k  + C_{ij}~,
\label{eqn:bdef}
\end{equation}
and the internal noise term,
\begin{equation}
C_{ij} \equiv (\mathbf{I}-\mathbf{S}~\mathbf{S}^\dagger)_{ij}~\frac{h\nu}{2}\frac{e^x+1}{e^x - 1}~.
\label{eqn:cdef}
\end{equation}
Here, $S_{ij}$ is the standard scattering matrix, which couples an
output amplitude to the inputs at each port of a network, as in Figure 
\ref{fig:smatrix2}. It is defined by $a_i = \sum_j
S_{ij} b_j$. In Equation \ref{eqn:cdef}, $x \equiv h\nu/kT_S$, where
$T_S$ is the thermodynamic temperature of the system $S$, and the
$n_k$ in Equation \ref{eqn:bdef} are the occupation numbers of the
modes at port $k$. We take ports 0 and 1 to label the two input
polarization states, and let ports 2 and 3 label the two bolometers in a PSB
pair.  For simplicity, we assume that $n_2 = n_3 = 0$ (\ie, the
detectors are extremely cold with respect to the background), and that
the input populations $n_0=n_1=n(\nu,T_{\mathrm{load}})$ imply no net
polarization in the background.

The internal noise term, $C_{ij}$ arises as a result of losses in the
system. Any mechanism causing loss implies a thermal noise
contribution, $c_i$, to the outgoing signals 
$$a_i = \Sigma_j S_{ij} b_j + c_i $$
which depends on the temperature of the lossy component.

In an ideal lossless network, the system\rq s thermal noise term will
vanish  since $S$ is unitary, $(I-S~S^\dagger)_{ij} = 0$.  In this
case, the only nonzero terms in the scattering matrix are
$S_{20}=S_{31}=1$.  Since the only nonzero populations are
$n_{0,1}=n$, the covariance matrix contains only terms with
$B_{20}=B_{31}=n\cdot h\nu$. Under this assumption the detectors\rq ~noise is
uncorrelated, and the autocorrelations satisfy
\begin{equation}
\sigma_{ii}^2 = \frac{(h\nu)^2}{\tau}\int d\nu n (n + 1).
\label{eqn:covar}
\end{equation}
Therefore the $1\sigma$ uncertainty in the incident power due to
intrinsic background fluctuations is
\begin{equation}
\sigma_{photon} = \frac{h\nu}{\eta}\sqrt{\frac{\Delta\nu}{\tau}}\sqrt{\eta
    n(\eta n+1)}~.
\label{eqn:bolo_phot_nep}
\end{equation}
Note that in the above we have explicitly included the optical efficiency,
$\eta \equiv |S_{20}|^2 = |S_{31}|^2$. This result differs from the
familiar Dicke radiometer equation describing coherent receivers,
\begin{equation}
\sigma_{photon} = \frac{h\nu}{\eta}\sqrt{\frac{\Delta\nu}{\tau}}(\eta n+1).
\label{eqn:dicke}
\end{equation}
While Equation \ref{eqn:dicke} is the limiting form of Equation
\ref{eqn:bolo_phot_nep} for large occupation number $n$, it is
instructive to derive Equation \ref{eqn:dicke} from Equation
\ref{eqn:photon_covar}.

\subsubsection*{Example 1: The Dicke radiometer equation}
The scattering matrix for an idealized coherent receiver with perfect
isolation contains a single nonzero term, $|S_{10}|^2 = G$, where $G$
is the gain of the system. An amplifier can be thought of as a
population characterized by an inverted distribution of energy levels,
such as that found in a maser or laser. Such systems are conveniently
described in terms of a negative temperature.
As $T\rightarrow ~ ^{-}0$, the sign of the $C_{ij}$ from
Equation \ref{eqn:cdef} is reversed.  The only nonzero element of
$\mathbf{C}$ is $C_{11}= G-1$, and therefore $B_{11}=Gn+ G-1$.  
Application of Equation \ref{eqn:photon_covar} gives
\begin{eqnarray}
\sigma^2_{11} &=&\frac{(h\nu)^2}{\tau}\int d\nu ~
 G^2\left[n+1-(G)^{-1}\right]\left[n+1\right] \\
&=& \Delta\nu ~ \frac{(h\nu)^2}{\tau} ~G^2 \left[ (n+1)^2 -
  \frac{(n+1)}{ G}\right].
\end{eqnarray}
In the limit that $G$ is significantly larger than unity, the second
term becomes negligible.  Referencing the noise to the input, we
recover the (lossless) Dicke radiometer equation, Equation \ref{eqn:dicke},
$$\sigma_{11} = h\nu\sqrt{\frac{\Delta \nu}{\tau}}(n+1).$$

\subsubsection*{Example 2: Polarization sensitive bolometers}

A dual polarized, single-moded receiver (coherent or bolometric) is completely
described by a four port network.  Polarization sensitive bolometers
and coherent receivers using orthogonal mode transducers (OMTs) are
two examples of such systems.  We now derive the photon noise
properties of a PSB pair.

The action of the network, $S$, is that of
an imperfect polarized beam splitter, with two inputs and two detectors.
The PSBs (two of the four ports, labeled say, as numbers 2 and 3) are
assumed to be at cryogenic temperatures, and therefore contribute
negligibly to the photon occupation number.  Therefore, the entries in the
scattering matrix relevant to the observed photon noise are limited to
the lower left quadrant, namely $S_{20}=\gamma, S_{31}=\gamma^\prime,
S_{21}=\delta$, and $S_{30}=\delta^\prime$.  Here the parameters
$\gamma$ and $\delta$ describe the efficiency of transmission of the
copolar amplitude and the crosspolar amplitude, respectively, and in
practice $\gamma >> \delta$.

Only the lower right quadrant of $SS^\dagger$ is nonzero,
\begin{equation}
\begin{array}{cc}
(SS^\dagger)_{22} = \gamma^2+\delta\delta^\prime &
(SS^\dagger)_{23} = \delta ~(\gamma+\gamma^\prime) \\  
(SS^\dagger)_{32} = \delta^\prime ~(\gamma+\gamma^\prime) &
(SS^\dagger)_{33} = \gamma^{\prime 2}+\delta\delta^\prime
\end{array}
\end{equation}
The power coupling terms, $B_{ij}$, of interest are given by
\begin{eqnarray}
B_{22} &=& \left(|S_{20}|^2 ~n_0 + |S_{21}|^2 ~n_1\right)~h\nu + C_{22} \\
&=& \left( \gamma^2 ~n_0 + \delta^2 ~n_1 +
[1-(\gamma^2+\delta\delta^\prime)]~ n_c\right)~h\nu \\
B_{23} &=& \left(S_{20}S_{30} ~n_0 + S_{21}S_{31} ~n_1\right)~h\nu + C_{22} \\
&=& \left(\gamma\delta^\prime ~n_0 + \gamma^\prime \delta ~n_1 -
\delta~(\gamma+\gamma^\prime)~ n_c \right)~h\nu
\end{eqnarray}
where we have written the thermal contribution of the network
$$ n_c \equiv \frac{1}{2}\frac{e^x+1}{e^x-1}. $$
In the case of PSBs, the source of the modal coupling are the
detectors themselves and/or the optics. In the case of the detectors,
they are extremely cold compared to the background.  For \boomn, the
reimaging optics and filters are also cooled, and have low
emissivity.  Therefore, we assume the thermal noise contribution of the
network, $n_c$, is very small compared to the background populations,
$n_i$. Furthermore,
we assume that the background is isotropic, \ie, $n_0 = n_1 = n$.
The covariance of the photon noise is then fully described by
\begin{eqnarray}
\sigma_{ii}^2 &=& (h\nu)^2 \frac{\Delta\nu}{\tau} \left[~
(\gamma^2+\delta^2)^2 ~n^2 +(\gamma^2 + \delta^2)
  ~n~\right]\label{eqn:phot_autocorr} \\
\sigma_{ij}^2 &=& (h\nu)^2 \frac{\Delta\nu}{\tau} \left[~ ( 2\gamma\delta
  )^2 ~ n^2 ~\right].
\label{eqn:bose_corr}
\end{eqnarray}
The autocorrelation, Equation \ref{eqn:phot_autocorr}, contains terms
proportional to both $n^2$ and $n$.  The former is commonly referred
to as the Bose contribution, or as a ``photon bunching'' term.
Equation \ref{eqn:bose_corr} shows that correlations between devices
are proportional only to the Bose contribution, implying that PSBs 
operating under higher background loading conditions will exhibit a
higher proportion of correlated noise than the same instrument
operating in a lower background.  For an idealized system,
in which the polarization leakage $\delta$ is zero, the covariance between
detectors vanishes since the two linear polarization states are statistically 
independent of one another.


In practice, we estimate the total optical background power, $Q$,
arising from the \cmbn, atmosphere, the telescope, and emission from
within the cryostat.  For simplicity, this optical background is
treated as having originated from a single thermal source at an
effective temperature $T_{RJ}=Q/\eta k_B\Delta\nu$.
The noise equivalent power from the
background fluctuations is then given by Equation \ref{eqn:phot_autocorr},
\begin{equation}
\mathrm{NEP}_{photon}^2 \simeq 2 h\nu ~Q ~\left(1+\eta ~n(T_{RJ})\right).\label{eqn:nep_phot}
\end{equation}
This is, of course, only approximate as we do not treat
the background sources independently.  It is often the case, however,
that a single thermal source contributes the majority of the
background optical power.


\section{The Jacobi method}\label{sec:jacobi}

We outline the application of the Jacobi method to the problem of mapmaking
from scanning experiments, much of which can be generalized to other
iterative algorithms such as the method of preconditioned conjugate
gradients.  We loosely follow the more complete discussions of the topic which
can be found, for example, in \citet{templates,acton,young}, and
\citet{numrec}.

The Jacobi method is a robust numerical method of solving a
set of linear equations, $\mathbf{Ax}=\mathbf{b}$, for which the
matrix $A$ is (or can be arranged to be) diagonally dominant.  
The great strength of the Jacobi method is that, subject to this requirement,
it is guaranteed to converge although it may do so relatively slowly.  
Given a trial solution, one may estimate a new solution
without inverting the matrix $\mathbf{A}$ simply by solving for each 
component, $x_i^{k+1}$, given an estimate of the values $\{x_i^k\}$, 
$$x_i^{k+1} = A_{ii}^{-1} \left( b_i -\sum_{j\ne i} A_{ij}
x_j^{k}\right).$$

It is often convenient, and advantageous from a numerical point of view,
to write the above in terms of a correction to the previous iteration,
$$x_i^{k+1} = x_i^{k}+\delta x_i^{k+1}$$
where
\begin{equation}
\delta x_i^{k+1} \equiv \eta ~ A_{ii}^{-1} \left( b_i -\sum_{j}
A_{ij}x_j^{k}\right). \label{eqn:jacobi}
\end{equation}
Here we have inserted a convergence parameter $\eta \lesssim 1$, which may
be tuned to aid the convergence of the algorithm.  In the limit that
$\mathbf{A}$ is diagonal, the optimal value is $\eta=1$.  Generally
speaking, the larger the off-diagonal terms become, the lower the
optimal value of $\eta$.  Clearly the diagonals of $\mathbf{A}$ must
not be near zero.  Furthermore, as can be seen from Equation
\ref{eqn:jacobi}, the solution will diverge if the absolute value of
the sum of the off-diagonals is greater than the diagonal element of
each row.

As an example, consider the following linear system:
\begin{equation}
\left(\begin{array}{ccc}
~5 & -2 & ~1 \\
~5 & -7 & ~1 \\
-2 & ~1 & ~6 \end{array}\right)
\mathbf{x}=\left(\begin{array}{c}
-1 \\ ~0 \\ ~1\end{array}\right)
\end{equation}
Setting $\eta=1$, and using the above procedure results in the
following sequence of solutions:
\begin{eqnarray*}
x_0 &=& (~0.000,~0.000,~0.000) \\
x_1 &=& (-0.200,~0.000,~0.167) \\
 &\vdots & \\
x_5 &=& (-0.291,-0.187,~0.102) \\
 &\vdots & \\
x_\infty &=& (-0.300,-0.200,~0.100)
\end{eqnarray*}

This example converges to twelve significant digits after 40
iterations, largely independent of $\mathbf{x}_0$, the trial solution.  The
rate of convergence does not scale strongly with the array size, so
the solution is an efficient way of solving large systems of equations.

A minor modification to the above procedure results in the
Gauss-Seidel algorithm, for which the estimate for each value
$x_i^{k+1}$ incorporates the most recent estimate of the parameters
$\{x_{j}^{k+1}\}_{j<i}$ instead of the set of values from the previous 
iteration.  This procedure is less numerically robust, but tends to
converge more rapidly than Jacobi iteration.

The application to Equation \ref{eqn:map} is clear;  the Jacobi method
provides a robust method of solving for the general least squares map.
Equating Equations \ref{eqn:map} and Equation \ref{eqn:jacobi} we find 
the correspondence,
\begin{eqnarray*}
\mathbf{A} &\rightarrow & \mathbf{C}_N^{-1} \equiv (\mathbf{A}^T \mathbf{N}^{-1} \mathbf{A} ) \\
\mathbf{x} &\rightarrow & \mathbf{\widetilde{m}} \\
\mathbf{b} &\rightarrow & \mathbf{A}^T \mathbf{N}^{-1} \mathbf{d}
\end{eqnarray*}
Recall that the matrix $\mathbf{A}$, which appears on the right hand side, is
the pointing matrix and should not be confused with the general
linear system described in Equation \ref{eqn:jacobi}.  The
algorithm we use for calculating the correction to an estimate of the
least squares map, $\mathbf{\widetilde{m}}^k$, is simply
$$ \delta \mathbf{\widetilde{m}}^{k+1} ~\propto ~\mathrm{diag}(\mathbf{A}^T
\mathbf{N}^{-1} \mathbf{A})^{-1} \mathbf{A}^T \mathbf{N}^{-1} (\mathbf{d}-\mathbf{A} \mathbf{\widetilde{m}}^k)~.$$
In practice, one can simultaneously solve for the noise covariance
matrix of the data, $\mathbf{N}$. This typically results in slightly
slower convergence of the algorithm than when using a fixed noise estimate.


\end{document}